\documentclass[aps,pra,twocolumn,superscriptaddress,nofootinbib,floatfix]{revtex4-2}

\usepackage[utf8]{inputenc}
\usepackage{amsmath,amssymb,amsfonts,amsthm}
\usepackage{mathrsfs}
\usepackage{dsfont}
\usepackage{empheq}
\usepackage{float}
\usepackage{braket}            % Dirac notation
\usepackage{bm}                % Bold math
\usepackage{graphicx}          % Figures
\usepackage[singlelinecheck=false,justification=raggedright]{subcaption}
\usepackage{hyperref}          % Links
\usepackage{mathtools}
\usepackage{microtype}
\usepackage{color}
\usepackage{enumitem}
\newtheorem{theorem}{Theorem}
\theoremstyle{remark}
\newtheorem{remark}{Remark}
\DeclarePairedDelimiter{\comm}{[}{]}
\DeclarePairedDelimiter{\norm}{\lVert}{\rVert}
\begin{document}

% -----------------------------------------
% Title & Author Block
% -----------------------------------------
\title{Open-System Adiabatic Quantum Search under Dephasing}

\author{Afaf El Kalai}
\email{afaf.el-kalai@tum.de}
\affiliation{Technical University of Munich, School of Natural Sciences, Department of Physics, Garching, Germany}
\affiliation{Siemens AG, Garching, Germany}

\author{Peter J.~Eder}
\email{peter-josef.eder@siemens.com}
\affiliation{Siemens AG, Garching, Germany}
\affiliation{Technical University of Munich, School of CIT, Department of Computer Science, Boltzmannstra{\ss}e 3, 85748 Garching, Germany}

\author{Christian B.~Mendl}
\email{christian.mendl@tum.de}
\affiliation{Technical University of Munich, School of CIT, Department of Computer Science, Boltzmannstra{\ss}e 3, 85748 Garching, Germany}
\affiliation{Technical University of Munich, Institute for Advanced Study, Lichtenbergstra{\ss}e 2a, 85748 Garching, Germany}

\date{7 April 2026}

% -----------------------------------------
% Abstract
% -----------------------------------------
\begin{abstract}
Adiabatic quantum algorithms must evolve slowly enough to suppress non-adiabatic transitions while remaining fast enough to be practical. In open systems, this trade-off is reshaped by decoherence. For Hamiltonians subject to dephasing Lindbladians, Avron et al. [\href{https://journals.aps.org/pra/abstract/10.1103/PhysRevA.82.040304}{Phys. Rev. A 82, 040304(R) (2010)}] showed that a unique timetable exists that maximizes the fidelity with a target state. This optimal schedule is characterized by a constant tunneling rate along the adiabatic path. In this work, we revisit their analysis and apply it to the adiabatic Grover search framework, obtaining closed-form expressions for the optimal evolution schedule, the minimum runtime, and the resulting achievable fidelity. Moreover, by invoking an energy--time uncertainty argument, we identify a critical dephasing threshold beyond which further noise-assisted acceleration is prohibited, showing that the apparent advantage of strong dephasing is a Zeno-type stabilization mechanism rather than a conventional algorithmic speedup over unitary Grover search.

\end{abstract}

\maketitle

% -----------------------------------------
% I. INTRODUCTION
% -----------------------------------------
\section{Introduction}

Identifying ground states of quantum many-body systems lies at the heart of quantum simulation, condensed matter theory, and optimization. Yet, even with major advances in classical techniques such as density functional theory, quantum Monte Carlo, or tensor networks, one cannot hope to develop a universal classically efficient method, since the general problem remains theoretically quantum Merlin Arthur (QMA)-hard~\cite{QMAhard}. Quantum algorithms promise a way forward, and among them, the quantum adiabatic paradigm occupies a special place: it replaces the discrete logic of the circuit model with a continuous interpolation between Hamiltonians whose ground states encode computational solutions. Since its introduction by Farhi \emph{et al.}~\cite{farhi} and the subsequent proof of equivalence with the circuit model~\cite{equivalent}, adiabatic computing has become a cornerstone of quantum algorithm design. Several works have suggested that its continuous dynamics may offer a degree of intrinsic robustness to certain noise processes~\cite{12,13}, with the possibility that environmental interactions could even enhance performance~\cite{14}.

However, scalable adiabatic computation faces a fundamental bottleneck: the shrinking spectral gap that typically accompanies increasing problem sizes. According to the adiabatic theorem, the runtime must grow at least as fast as the inverse square of the minimum gap, which for many Hamiltonians decreases  exponentially~\cite{vanDamMoscaVazirani2001, AltshulerKroviRoland2010}. This sensitivity becomes especially acute near quantum phase transitions where diabatic excitations proliferate~\cite{Schuetzhold2006}. At the same time, realistic devices are never perfectly isolated; environmental noise acts throughout the evolution, compounding the difficulty of maintaining adiabaticity. Efforts to circumvent these limitations have inspired a variety of shortcuts to adiabaticity ~\cite{shortcuts}, most notably transitionless quantum driving ~\cite{Demirplak2003}, which cancels diabatic transitions via auxiliary control fields, as well as Zeno-~\cite{Facchi2002,zeno} and dissipation-~\cite{Verstraete2009,Peter} based search protocols that exploit strong measurements or engineered Lindbladians to drive the system toward the ground state.

Grover’s algorithm, which searches for a marked item in an unstructured search space of size $N$, provides a particularly clear setting in which to examine these issues~\cite{grover, farhi}. Due to its symmetry, its effective two-level structure enables analytical access, and its optimal quadratic speedup highlights the power of adiabatic computation in the ideal, closed-system case. At the same time, the model features a vanishing minimum gap $g_{\min}\sim N^{-1/2}$, making it sensitive to diabatic transitions~\cite{rolandcerf}. For this reason, it has become a standard benchmark for understanding how realistic noise sources shape the performance and limitations of adiabatic algorithms \cite{albash2015decoherence}.

Among such noise processes, we focus on energy-basis dephasing. It preserves energy populations while suppressing coherences and, within the framework of the open-system adiabatic theorem~\cite{SarandyLidar2005}, the evolution of its instantaneous steady states is governed by a geometric transport law. In this case, the stationary manifold coincides with the Hamiltonian eigenbasis, and slow driving induces a parallel transport within this manifold -- defining the ideal adiabatic trajectory -- with deviations that scale quadratically in the interpolation speed~\cite{AvronFraas, AdiabaticResponse}. As we show in this work, Grover search illustrates this structure clearly: when noise respects the system's spectral geometry, it can, in certain regimes, stabilize and guide the adiabatic trajectory rather than merely degrade it.

Of the many strategies proposed to accelerate or stabilize adiabatic evolution, in this paper we focus on optimizing the interpolation schedule, often referred to as the \emph{quantum adiabatic brachistochrone}~\cite{brachistochrone}. In the specific context of unstructured search, Roland and Cerf~\cite{rolandcerf} identified an optimal schedule for adiabatic Grover search that slows near the minimum gap and recovers the characteristic $\mathcal O(\sqrt{N})$ speedup of the circuit-model algorithm. Extending these ideas to open quantum systems, Avron \emph{et al.}~\cite{avron} analyzed dephasing Lindblad dynamics and showed that the unique optimal schedule maintains a constant instantaneous tunnelling rate, while King \emph{et al.}~\cite{trajectories} constructed the adiabatic optimal schedule for arbitrary two-level systems governed by a dephasing Lindblad master equation. Related Grover-type speedups arising from Zeno mechanisms have also been explored in different settings, including repeated-measurement and decoherence-induced regimes~\cite{zeno}. Building on this line of work, we study adiabatic Grover search in the presence of energy-basis dephasing.  In particular, we:

\begin{itemize}
\item revisit the analysis of Avron \emph{et al.}~\cite{avron} to obtain the optimal interpolation schedule, minimal runtime, and resulting tunneling probability (infidelity) for the Grover problem;
\item compare the simulation times of the unitary adiabatic Grover algorithm using the Roland--Cerf schedule with dephasing dynamics under the schedule derived here;
\item use an energy--time uncertainty argument to identify a critical dephasing rate beyond which further noise-assisted acceleration is not possible for a universal bath.  We conclude that dephasing can assist by stabilizing a target eigenspace in the Zeno regime, but it should not be interpreted as a generic algorithmic improvement over the unitary Grover scaling for an unstructured search.
\end{itemize}

\section{Adiabatic evolution in closed and open systems.}  
\label{sec:adiabatic_evolution}

In the ideal unitary setting, the system evolves according to the Schrödinger equation and the adiabatic theorem ensures that a system initialized in the ground state of the initial Hamiltonian \(H(0)\) remains close to the instantaneous ground state of \(H(t)\), provided the Hamiltonian varies sufficiently slowly. In many systems, this condition is well approximated by requiring the total runtime $T$ to scale inversely with the square of the minimum spectral gap \(g_{\min}\). This condition can be quantitatively expressed as
\begin{equation}
T \gg \max_t \frac{\norm{\dot{H}(t)}}{g_{\min}^2}.
\end{equation}
More rigorous versions can be found for example in Ref.~\cite{stateart}.

In realistic scenarios, quantum systems are never fully isolated but interact with external environments, leading to non-unitary dynamics. The state of such a system, described by the density operator \(\rho(t)\), evolves according to the Lindblad master equation
\begin{equation}
\label{eq:masterequation}
\frac{d}{dt}\rho(t) = \mathcal{L}(t)\rho(t),
\end{equation}
where $\mathcal{L}(t)$ is a time-dependent \emph{Liouvillian} superoperator. The Lindblad equation rigorously describes Markovian (i.e., memoryless) dynamics in the weak system--environment coupling limit~\cite{weakcoupling}. We restrict our analysis to finite-dimensional Hilbert spaces $\mathcal{H}$, allowing the time evolution of $\rho(t)$ to be described by a \emph{quantum dynamical semigroup}. Under these conditions, the generator $\mathcal{L}(t)$ takes the \emph{Gorini-Kossakowski-Sudarshan-Lindblad} (GKSL) form
\begin{align}
\mathcal{L}(t)\rho(t) &= -i[H(t), \rho(t)] \label{eq:Lindblad} \\
&\quad + \sum_j \left( 2 L_j(t) \rho(t) L_j^{\dagger}(t) - \{L_j^{\dagger}(t) L_j(t), \rho(t)\} \right), \notag
\end{align}
where \(\{A,B\}=AB+BA\) denotes the anticommutator.  
Here, \(H(t)\) is the effective system Hamiltonian, potentially including environment-induced corrections beyond the closed-system term, while \(L_j(t)\) are Lindblad (jump) operators modeling dissipative channels. This structure ensures that the evolution of $\rho(t)$ is governed by a \emph{completely positive, trace-preserving} (CPTP) map, which preserves the physical properties of the density matrix throughout its evolution.

In the more general open-system setting, the adiabatic theorem can be reformulated in terms of the instantaneous stationary manifold of $\mathcal{L}(t)$. Let $\mathcal{P}(t)$ denote the projection onto this manifold. The \emph{parallel transport} operator $\mathcal{T}(s,s')$, defined by 
\begin{equation}
\frac{d}{ds}\mathcal{T}(s,s') = \comm*{\dot{\mathcal{P}}(s),\mathcal{P}(s)}\,\mathcal{T}(s,s'), \qquad \mathcal{T}(s',s') = \mathds{1},
\end{equation}
describes the ideal adiabatic path that keeps the system confined within the instantaneous steady-state subspace \cite{AvronFraas}. Here, $s$ and $s'$ denote normalized time parameters, with $s = t/T$ and $0\leq s\leq1$. The condition $\mathcal{T}(s',s') = \mathds{1}$ ensures that at the reference time $s'$ the operator acts as the identity, leaving the initial state unchanged. Physically, $\mathcal{T}(s,s')$ generalizes Berry’s geometric transport~\cite{Berry84} to open systems and defines the ideal adiabatic trajectory along which the system would evolve in the adiabatic limit. In contrast, the actual evolution of the system, including all nonadiabatic effects, is described by the CPTP map $\mathcal{U}(s,0)$. Avron et al.~\cite{AvronFraas} showed that the deviation between $\mathcal{U}(s,0)$ and the ideal adiabatic evolution $\mathcal{T}(s,0)$ is bounded by
\begin{equation} 
\label{eq:openQAT} 
\norm*{\left(\mathcal{U}(s,0)-\mathcal{T}(s,0)\right)\,\mathcal{P}(0)} \le \frac{C}{T}, 
\end{equation} 
where the constant \(C\) was later determined by Venuti et al.~\cite{Venuti2016Adiabaticity} as
\begin{align} 
\label{eq:adiabaticTheoremC} C &= \|\mathcal{S}(s)\|\|\mathcal{P}'(s)\| + \|\mathcal{S}(0)\|\|\mathcal{P}'(0)\| \notag \\ &\quad + \int_{0}^{s} \left\|\left[\mathcal{S}'\mathcal{P}' + \mathcal{S}\mathcal{P}''\right](\sigma)\right\|\,d\sigma. 
\end{align}
\noindent
Inequality \eqref{eq:openQAT} constitutes the quantum adiabatic theorem for open systems. For an initial steady state
\(\tilde{\rho}(0)\in\operatorname{Ker}\mathcal{L}(0)\), the ideal adiabatic evolution
\(\tilde{\rho}(s)=\mathcal{T}(s,0)\tilde{\rho}(0)\)
remains in \(\operatorname{Ker}\mathcal{L}(s)\).
Under the actual dynamics \(\rho(s)=\mathcal{U}(s,0)\tilde{\rho}(0)\), the deviation obeys
\[
\|\rho(s)-\tilde{\rho}(s)\|_1
\le
\|[\mathcal{U}(s,0)-\mathcal{T}(s,0)]\mathcal{P}(0)\|
\le
\frac{C}{T}.
\]
Hence, if \(T \ge C/\eta\) for a target precision \(\eta>0\),
the evolved state remains within distance \(\eta\) (in trace norm)
of the instantaneous steady state at all times. In the following, by evaluating $C$ as a function of system size $N$, we will determine the scaling of the total runtime $T$ with $N$ for adiabatic Grover search under dephasing.

\section{Dephasing Dynamics in Open Quantum Systems}
\label{section:lindblad}
Building on the general framework outlined above, we now focus on \emph{dephasing Lindbladians}, a subclass of GKSL generators describing energy-preserving decoherence. Following the definition of Avron \emph{et al.}~\cite{AvronFraas}, a generator \( \mathcal L\) is dephasing with respect to a Hamiltonian \(H\) if
\(\ker \mathcal L^* \supset \ker[H,\cdot],\)
which is equivalent to taking jump operators that are functions of the Hamiltonian and hence commute with it,\([L_j,H]=0\).

Writing \(H\) in its spectral decomposition \(H = \sum_a E_a P_a\), with orthogonal projectors \(P_a\), one obtains the general \textit{dephasing Lindbladian} (see Appendix~\ref{app:gksl} for the derivation)

\begin{equation}
\label{eq:general_depha}
\mathcal{L}(\rho)
= -i[H, \rho]
+ 2\!\sum_{a,b}\! \gamma_{ab} P_a\rho P_b
- \sum_a \gamma_{aa}\{P_a,\rho\},
\end{equation}
where the positive-semidefinite matrix \(\gamma_{ab}\) encodes the dephasing rates induced by the bath. The particularly symmetric case \(\gamma_{ab} \rightarrow \gamma\,\delta_{ab}\) corresponds to uniform dephasing across all levels, reducing Eq.~\eqref{eq:general_depha} to the simple form:
\begin{equation}
\label{eq:dephasing}
\mathcal{L}(\rho) = -i[H, \rho] - \gamma \sum_{a \neq b} P_a\rho P_b,
\end{equation}
which preserves populations \(p_a = \operatorname{tr}(P_a\rho)\) while exponentially damping coherences with rate \(\gamma\).  
Because \(\operatorname{tr}[H\,\mathcal{L}(\rho)] = 0\), energy is conserved, and any state diagonal in the energy basis is stationary. 

Equation~\eqref{eq:dephasing} should therefore be understood as pure energy-basis dephasing rather than relaxation: it does not drive an arbitrary state to the ground state, but suppresses coherences between instantaneous eigenspaces and can thereby stabilize an adiabatic trajectory initialized in the ground-state manifold. Physically, this model may be interpreted as continuous weak measurement of the instantaneous energy. In the Grover setting, such a coupling requires access to the oracle-defined Hamiltonian structure, but not necessarily classical knowledge of the marked state. A physical realization leading to this type of dephasing dynamics was discussed in Ref.~\cite{trajectories}.

\medskip
\noindent

We now apply this framework to the adiabatic interpolation between an initial Hamiltonian \(H_0\), with an easy-to-prepare ground state, and a target Hamiltonian \(H_1\). Let \(H_q\), \(q \in [0,1]\), be a path in the space of Hamiltonians, e.g.,
a linear interpolation,
\begin{equation}
\label{eq:linearinterpolation}
    H_q=(1-q) H_0+q H_1, \quad q \in[0,1].
\end{equation}
Further, assume that at each value of $q, H_q$ has a non-degenerate spectrum
\begin{equation}
\label{eq:spectral}
    H_q=\sum_{a=0}^{N-1} E_a(q) P_a(q),
\end{equation}
with ordered eigenvalues $E_0(q)<E_1(q)<\cdots<E_{N-1}(q)$ and associated spectral projections $P_a(q)$. Instead of varying $q$ at a constant rate, we allow an arbitrary smooth schedule $q=q(s)$. Here, $s=\varepsilon t$ rescales physical time $t$ by the adiabaticity parameter $\varepsilon=1 / T$, so that the full evolution takes place over a run-time $T$. The system, initially prepared in the ground state $P_0(0)$ of $H_0$, then evolves under the dephasing Lindbladian $\mathcal{L}_{q(s)}:=\mathcal{L}_{q}$ according to
\begin{equation}
\label{eq:Lindbladian}
    \varepsilon \dot{\rho}(s)=\mathcal{L}_{q}(\rho(s)) .
\end{equation}
Here and in the following, dots will denote derivatives with respect to $s$ and primes will denote derivatives with respect to $q$. Because $\mathcal{L}_q$ preserves populations in the instantaneous energy basis but damps coherences at rates determined by its positive-semidefinite rate matrix, the choice of schedule $q(s)$ directly controls how much leakage -- tunneling to excited states -- accumulates by the end of the run. We quantify this leakage by the cost functional
\begin{equation}
\label{eq:tunneling1}
    \mathscr{T}_{q, \varepsilon}(1)=1-\operatorname{tr}\left[P_0 \rho_{q, \varepsilon}(1)\right],
\end{equation}
and our goal is to find $q(s)$ that, subject to $q(0)=0$ and $q(1)=1$, minimizes $\mathscr{T}_{q, \varepsilon}(1)$ in the adiabatic limit $\varepsilon \rightarrow 0$. The leading-order behavior of this tunneling functional in the adiabatic limit is given by the following theorem due to Avron and Fraas~\cite{avron}.
\begin{theorem}[Avron–Fraas \cite{avron}]
\itshape
\label{thm:AF}
Let \(\mathcal L_q\) be the dephasing Lindbladian in Eq.~\eqref{eq:dephasing} 
and let \(\rho_{q,\varepsilon}(s)\) be the solution to Eq.~\eqref{eq:Lindbladian}. Under the gap condition \(E_a(q)\neq E_b(q)\) for all \(a\neq b\) and all \(q\in[0,1]\), the tunneling probability~\eqref{eq:tunneling1} at \(s=1\) satisfies
\begin{equation}
\label{eq:tunneling_main}
    \mathscr{T}_{q,\varepsilon}(1)
   \;=\;
 2\,\varepsilon
 \int_{0}^{1} 
      M(q)\,
      \dot q(u)^{\,2}\,\text{d}u
 \;+\;
 \mathcal O(\varepsilon^{2}),
\end{equation}
where the \emph{$q$‑dependent mass}
\begin{equation}
\label{eq:mass}
    M(q)
   \;=\;
 \sum_{a\neq 0}
   \frac{\gamma(q)\;
         \mathrm{tr}\!\bigl(P_a(q)\,{P_0'(q)}^{2}\bigr)}
        {[\,E_0(q)-E_a(q)\,]^{2}+\gamma(q)^{2}}
 \;\;\ge 0
\end{equation}
is independent of the particular parametrisation \(s\).
\end{theorem}

\noindent
This result shows that, under dephasing, the leading-order tunneling probability is a quadratic functional of the schedule velocity $\dot q$ and can be minimized by a suitable parametrization. Avron et al.~\cite{avron} derived the resulting optimal velocity and minimal tunneling bound. To make the optimization structure explicit, we briefly restate the variational formulation leading to their result. One considers the functional
\begin{equation}
S[q] = \int_0^1 M(q)\,\dot{q}^{2}\,ds,
\end{equation}
with boundary conditions \(q(0)=0\) and \(q(1)=1\). The Euler–Lagrange equation for \(f(q,\dot{q})=M(q)\dot{q}^{2}\) is
\begin{equation}
\frac{d}{ds}\!\left(2M(q)\dot{q}\right) - M'(q)\dot{q}^{2}=0,
\end{equation}
or equivalently,
\begin{equation}
\ddot{q} + \frac{M'(q)}{2M(q)}\dot{q}^{2} = 0.
\end{equation}
Setting \(P=\dot q\) and integrating gives
\begin{equation}
P\,\sqrt{M(q)} = \text{const} = \tau,
\end{equation}
which yields the optimal velocity profile
\begin{equation}
\label{eq:optimal-speed}
\boxed{\;
\dot q_{\mathrm{opt}}(s)
= \frac{\tau}{\sqrt{M(q_{\mathrm{opt}}(s))}}.
\;}
\end{equation}
The normalization constant is fixed by the boundary conditions as~\cite{avron}
\begin{equation}
\label{eq:tau}
    \tau = \int_0^1 \sqrt{M(q)}\,\text{d}q.
\end{equation}

Along the minimizing trajectory the quantity \(E := M(q)\dot{q}^{2}\) is constant, since the effective Lagrangian is independent of the reduced time~$s$. Consequently, the protocol slows down where \(M(q)\) is large -- typically near small gaps -- and speeds up where it is small, keeping the instantaneous leakage rate uniform. The corresponding minimal tunneling probability~\cite{avron}
\begin{equation}
\label{eq:tunneling_min}
\mathscr{T}_{\min}
= 2\varepsilon\,\tau^{2}
+ O(\varepsilon^{2}).
\end{equation}
\noindent
Because the integrand depends only on the instantaneous values of \(q\) and \(\dot q\), the optimal schedule is time-local, and convexity guarantees a \textit{unique minimizer} of the boundary-value problem~\cite{avron}. Integrating
\(
\mathrm{d}s = \sqrt{M(q)}/\tau\,\mathrm{d}q
\)
gives the parametrization
\begin{equation}
\label{eq:optimal-schedule-revised}
s(q) = \frac{1}{\tau}\int_0^{q}\sqrt{M(u)}\,\text{d}u,
\end{equation}
whose inverse defines the optimal schedule.

\section{Application to the Adiabatic Grover Search}
\label{sec:grover}
\subsection{Adiabatic Grover Problem in Unitary Dynamics}
\label{sec:unitary_grover}
The Grover oracle problem is defined by a function \( f : \{0,1\}^n \to \{0,1\} \) with \(N=2^n\), promised to satisfy \(f(x)=1\) for \(x \in \mathcal{M}\subset\{0,1\}^n\) and \(f(x)=0\) otherwise. For simplicity, we restrict ourselves to a single marked item, \(\mathcal{M}=\{m\}\). The goal is to identify \(m\) using as few oracle queries as possible~\cite{NielsenChuang2000}.

\noindent
Classical algorithms for exhaustive unstructured search exhibit an average query complexity that scales linearly with the system size $N$. In contrast, AQC encodes the problem in the ground state of the problem Hamiltonian
\begin{equation}
H_p = \mathds{1} - |m\rangle\langle m|,
\end{equation}
whose unique ground state corresponds to the marked item \(|m\rangle\), while all other computational basis states are degenerate with energy~1. The computation begins in the ground state  \(|\psi_0\rangle \) of the initial mixer Hamiltonian
\begin{equation}
H_0 = \mathds{1} - |\psi_0\rangle\langle \psi_0|,
\quad
|\psi_0\rangle = \frac{1}{\sqrt{N}}\sum_{i=0}^{N-1}|i\rangle.
\end{equation}

\noindent
Within the adiabatic framework of Section~\ref{section:lindblad}, we use the linear interpolation~\eqref{eq:linearinterpolation} with schedule \(q(s)\) and slow time \(s=\varepsilon t\). By symmetry, the Grover Hamiltonian can be written in the two-dimensional subspace spanned by \(|m\rangle\) and
\(
|m^{\perp}\rangle = \frac{1}{\sqrt{N-1}}\sum_{i\neq m}|i\rangle.
\)
In this basis,
\begin{equation}
\label{eq:groverHam}
H(q) =
\begin{pmatrix}
(1-q)\tfrac{N-1}{N} & -(1-q)\tfrac{\sqrt{N-1}}{N} \\
-(1-q)\tfrac{\sqrt{N-1}}{N} & q + (1-q)\tfrac{1}{N}
\end{pmatrix},
\end{equation}
with instantaneous eigenvalues \(E_{0,1}(q)=\tfrac{1}{2}(1\mp g(q))\) and gap
\begin{equation}
\label{eq:grover_gap}
g(q)=\sqrt{(1-2q)^2+\tfrac{4q(1-q)}{N}},
\end{equation}
which reaches its minimum value \(g_{\min}=1/\sqrt{N}\) at \(q=\tfrac{1}{2}\), signaling an avoided level crossing whose width shrinks as \(N^{-1/2}\). For a linear schedule \(q(s)=s\), the dynamics near this avoided crossing are well described by the Landau--Zener model~\cite{landauzener}, yielding an excitation probability
\begin{equation}
\label{eq:landau-zener}
\mathscr{T}_{\mathrm{LZ}}
\simeq \exp\!\Bigl[-\tfrac{\pi g_{\min}^2}{2\varepsilon}\Bigr].
\end{equation}

\begin{figure*}[!t]
    \centering

    % Top row: two figures side-by-side
    \begin{subfigure}[t]{0.48\textwidth}
        \centering
        \includegraphics[width=\textwidth]{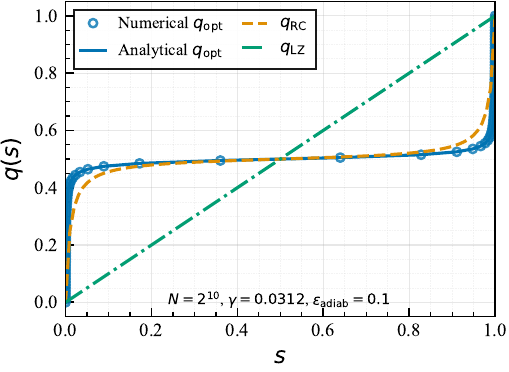}
        \caption{The optimal schedule \(q_{\rm opt}\) (Eq.~\eqref{eq:optimal_schedule}) for dephasing with \(\gamma = g_{\min}\) is shown, with circles marking the numerical solution to Eq.~\eqref{eq:grover-opt-speed}. Also shown: Roland–Cerf (RC) (Eq.~\eqref{eq:RCschedule}) and Landau–Zener (LZ) schedules.
        }
        \label{fig:1a}  
    \end{subfigure}
    \hfill
    \begin{subfigure}[t]{0.48\textwidth}
        \centering
        \includegraphics[width=\textwidth]{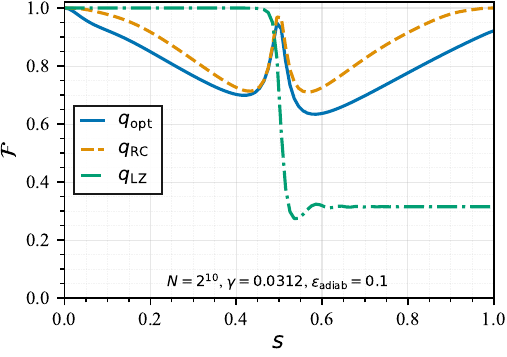}
        \caption{Instantaneous ground-state fidelity \(F(s)= \bra{\psi_{g}(s)}\rho(s)\ket{\psi_{g}(s)}\) is plotted against the dimensionless time \(s\) for the (i) dephasing-optimal Lindbladian (\(\gamma=g_{\min}\)), (ii) Roland–Cerf (RC), and (iii) Landau–Zener (LZ) protocols.    
        }
        \label{fig:1b}
    \end{subfigure}

    \vspace{1em} % Space between rows

    % Bottom row: wide figure
    \begin{subfigure}[t]{1.12\textwidth}
        \centering
        \includegraphics[width=\textwidth]{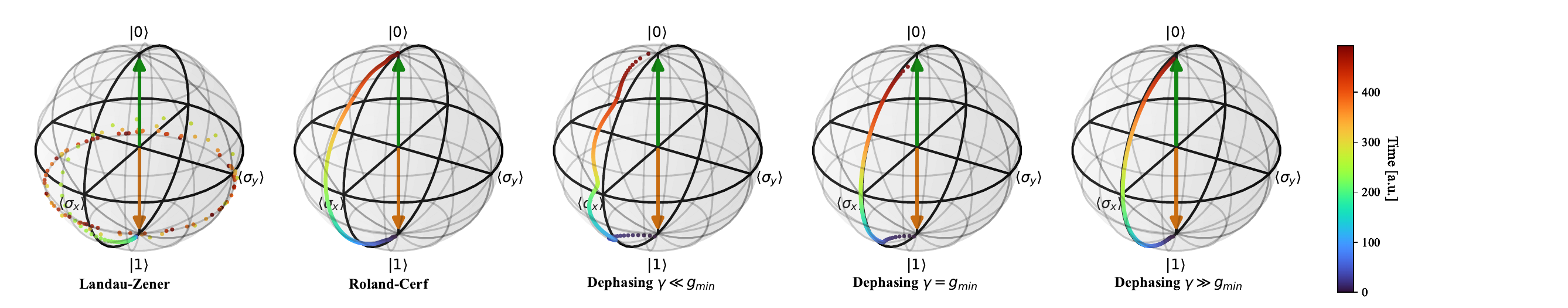}
        \captionsetup{width=\textwidth}
        \caption{
        Bloch sphere trajectories for unitary Landau–Zener transition, unitary Roland–Cerf adiabatic schedule, and Lindbladian \\ dynamics with an optimal schedule at three dephasing rates ($0.1 \, g_{\min}, g_{\min}, 10 \,g_{\min}$). Lindblad equations were solved \\ numerically using \texttt{solve\_ivp} with the high-order \texttt{DOP853} Runge–Kutta integrator.
        }
        \label{fig:1c}
    \end{subfigure}

    \caption{Dynamics of a 10-qubit adiabatic Grover problem under different control protocols for a fixed total evolution time \(T=T_{\textbf{RC}}\) as defined in Eq.~\eqref{eq:tRC} (with \( c=\varepsilon_{\rm adiab}\)).}
    \label{fig:combined}
\end{figure*}

As a consequence, achieving a fixed target tunneling probability requires a runtime scaling as \(T=O(g_{\min}^{-2})\), which for the Grover problem implies \(T=O(N)\). Thus, no linear (constant-rate) schedule can realize the adiabatic trajectory in sublinear time. To overcome this limitation, Roland and Cerf~\cite{rolandcerf} proposed a locally adiabatic schedule that adapts the evolution rate to the instantaneous spectral gap, \(\frac{dq}{dt} = c\,g^2(q),\) where the normalization constant \(c\) is fixed by the boundary conditions \(q(0)=0\) and \(q(T)=1\). Solving this equation yields the analytic schedule
\begin{equation}
\label{eq:RCschedule}
q_{\mathrm{RC}}(s)=\frac{1}{2}
+\frac{1}{2\sqrt{N-1}}
\tan\!\bigl[(2s-1)\tan^{-1}\!\sqrt{N-1}\bigr].
\end{equation}
The corresponding runtime
\begin{equation}
\label{eq:tRC}
T_{\mathrm{RC}} = \frac{N}{c\sqrt{N-1}}\tan^{-1}\!\sqrt{N-1}
\end{equation}
restores Grover’s quadratic scaling, proven to be optimal in scaling~\cite{Farhi1998}. Although this protocol achieves the asymptotic  \(\mathcal{O}(\sqrt{N})\) scaling, it does not saturate the fundamental \textit{quantum speed limit} (QSL) along the same adiabatic path (see Appendix~\ref{app:QSLproof}).
By contrast, \textit{counterdiabatic} quantum driving -- originally introduced by Demirplak and Rice~\cite{Demirplak2003} and Berry~\cite{Berry2009} -- can in principle saturate the QSL, a result later confirmed experimentally by Bason \textit{et al.}~\cite{Bason2012}.

\subsection{Adiabatic Grover Problem under Dephasing}
\subsubsection{Optimal schedule}
Next, we apply the Lindblad framework of Section~\ref{section:lindblad} to the adiabatic Grover search. In the effective two-level subspace, we re-derive the Bloch sphere velocity (also reported in Avron et al.~\cite{avron}) and solve equation~\eqref{eq:optimal-speed} to obtain the explicit optimal schedule. Writing the Hamiltonian in traceless form, it reads
\[
\tilde H_q = H_q - \tfrac{1}{2}\mathrm{Tr}[H_q]\mathds{1}
           = \tfrac12\,\vec r(q)\!\cdot\!\vec\sigma ,
\]
with Bloch vector components
\begin{equation}
\label{eq:B1}
r_x = 2(1-q)\sqrt{\tfrac{N-1}{N}}, \quad
r_y = 0, \quad
r_z = \tfrac{N-2}{N} - 2q\tfrac{N-1}{N}. 
\end{equation}
Using \((\vec r\!\cdot\!\vec\sigma)^2 = |\vec r|^2\mathds{1}\), the eigenvalues are \(\pm|\vec r(q)|/2\), so the instantaneous gap is \(g(q) = |\vec r(q)|\). Differentiating Eq.~\eqref{eq:B1} shows that \(\vec r(q)\) moves at constant speed in the $x$–$z$ plane; for the normalized vector \(\vec r_{\text{unit}}(q) = \vec r(q)/g(q)\), the angular velocity is
\begin{equation}
\label{eq:velocity}
v(q) = \|\vec r'_{\text{unit}}(q)\|
      = \frac{2}{g^2(q)}\sqrt{\frac{1}{N}-\frac{1}{N^2}},
\end{equation}
hence \(v(q)\propto g^{-2}(q)\), with the fastest rotation near \(q=\tfrac12\). For this two-level system, the mass term \(M(q)\) from Eq.~\eqref{eq:mass} evaluates to
\begin{equation}
\label{eq:2D-mass}
M(q) = \frac{\gamma(q)}{4}\,\frac{v^2(q)}{g^2(q)+\gamma^2(q)},
\end{equation}
giving the Grover-specific form
\begin{equation}
\label{eq:grover-mass}
M(q) = \frac{\gamma(q)(N-1)}{N^2}\,
       \frac{1}{g^4(q)\,[\,g^2(q)+\gamma^2(q)\,]}.
\end{equation}
Now inserting \(M(q)\) into the variational condition~\eqref{eq:optimal-speed} yields the differential equation for the optimal speed along the adiabatic path,
\begin{equation}
\label{eq:grover-opt-speed}
\dot q(s) = \frac{\tau N}{\sqrt{\gamma(N-1)}}\,
            g^2(q)\,\sqrt{g^2(q)+\gamma^2(q)},
\end{equation}
with \(\tau\) fixed by the boundary condition \(q(0)=0\) and \(q(1)=1\). Integration leads to the explicit optimal schedule \(q_{\mathrm{opt}}(s)\) (see the full derivation in Appendix~\ref{app:grover-schedule}):

\[
\boxed{%
  \begin{array}{rl}
    q_{\mathrm{opt}}(s) 
    &= \dfrac{1}{2} + \dfrac{1}{2} \,
      \dfrac{\sin\theta\; \sqrt{N\gamma^{2} + 1}}
           {\sqrt{(N - 1)\!\left[ N\gamma^{2} - (1 + N\gamma^{2}) \sin^{2}\theta \right]}},%
  \end{array}
}
\]
\vspace{-2ex}
\begin{equation}
\label{eq:optimal_schedule}
\theta 
= (2s - 1)\,
  \sin^{-1} \!\left( \frac{\gamma \sqrt{N - 1}}
                          {\sqrt{N\gamma^{2} + 1}} \right).
\end{equation}
Figure~\ref{fig:combined}a shows that the dephasing-optimal schedule closely resembles the Roland–Cerf schedule: in both, the evolution slows near the minimum gap and accelerates elsewhere.
\subsubsection{Optimal Tunneling Rate and Simulation Time}
Next, we quantify the performance of the optimal schedule by evaluating the final tunneling probability \(\mathscr{T}_{q,\varepsilon}(1)\) defined in Eq.~\eqref{eq:tunneling_main}, which measures leakage into the excited state manifold at the end of the evolution. For the dephasing-optimized schedule, we substitute the constant timescale $\tau$ into Eq.~\eqref{eq:tunneling_min}. The constant $\tau$ was computed using \textit{Mathematica} by evaluating the integral in Eq.~\eqref{eq:tau}, which is equivalent to the analytical expression derived in Eq.~\eqref{eq:constant} of Appendix \ref{app:grover-schedule}. This yields the analytical minimum

\begin{multline}
\label{eq:Imin}
\mathscr{T}_{\min} = \frac{2}{\gamma T}
\biggl[
  \arctan\left(\frac{1}{\gamma\sqrt{N}}\right) + \\
  \arctan\left(
    \frac{\sqrt{(1+\gamma^{2})(N-1)} - \sqrt{N}}{\gamma}
  \right)
\biggr]^{2}.
\end{multline}
Conversely, for a target tunneling probability \(\mathscr{T}\), the minimal simulation time consistent with this precision is obtained by inverting Eq.~\eqref{eq:Imin}, noting that the adiabaticity parameter is $\varepsilon=1/T$, 
\begin{multline}
\label{eq:Tmin}
T_{\min}= \frac{2 \tau}{\mathscr{T}} = \frac{2}{\gamma\,\mathscr{T}}
\biggl[
  \arctan\!\!\left(\frac{1}{\gamma\sqrt{N}}\right) + \\
  \arctan\left(
    \frac{\sqrt{(1+\gamma^{2})(N-1)} - \sqrt{N}}{\gamma}
  \right)
\biggr]^{2}
\end{multline}
\begin{figure}[!t]
  \centering
  \includegraphics[width=1.1\columnwidth]{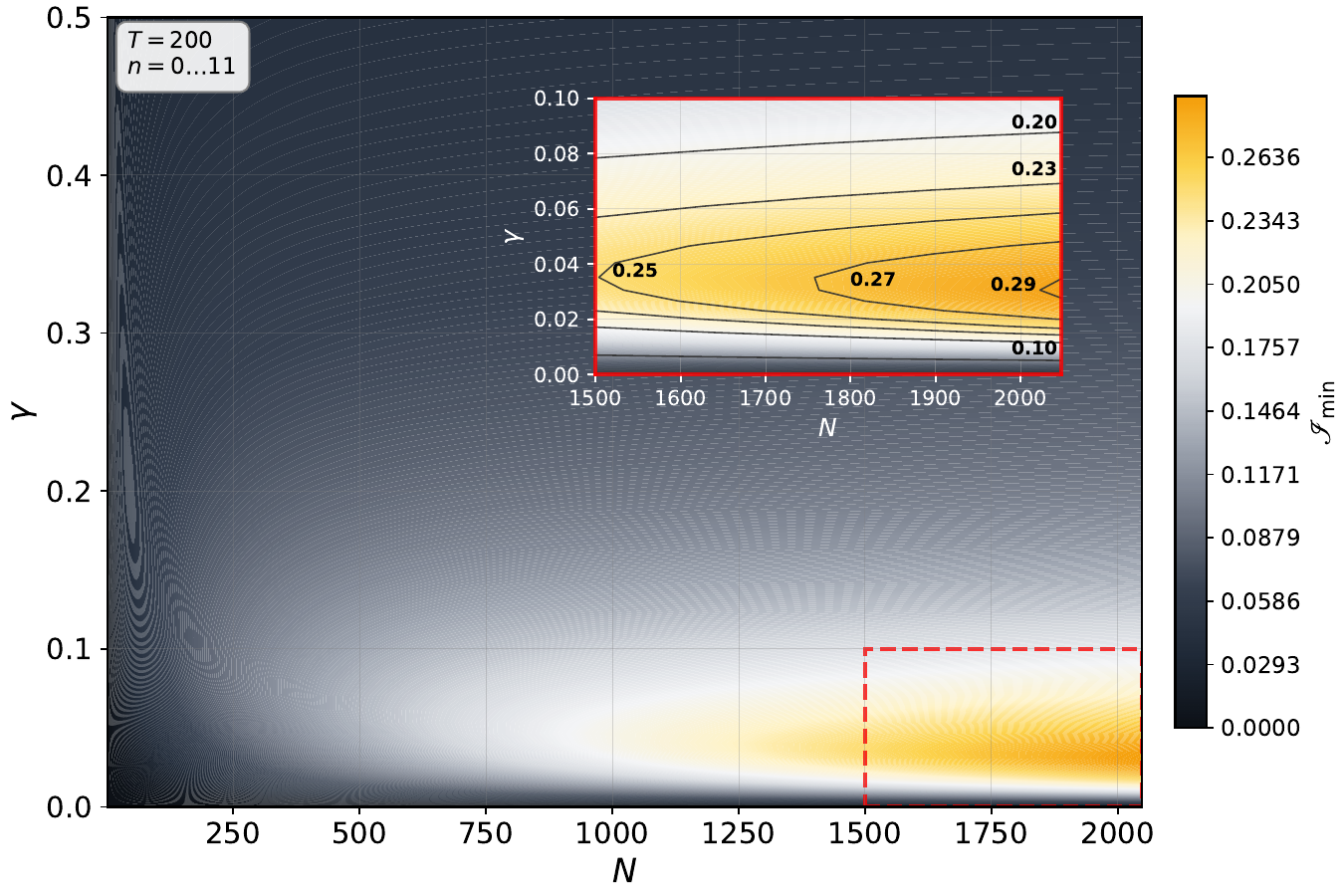}
  \caption {Contour plot of the minimum tunneling probability $\mathscr{T}_{\min}$ of Eq.~\eqref{eq:Imin} versus the system size and dephasing strength for a fixed simulation time $T=200$, in the spirit of Ref.~\cite{trajectories}. System sizes are \(N = 2^n\) up to 11 qubits. Small values of $\mathscr{T}_{\min}$ correspond to a large overlap with the ground state.}
  \label{fig:contour_plot}
\end{figure}
These expressions reveal the trade-off between dephasing strength $\gamma$, spectral gap $g_{\text{min}}$, and runtime $T$: a smaller minimum gap (or larger system size) demands either stronger dephasing or longer simulation times to maintain a fixed accuracy.  Figure~\ref{fig:contour_plot} shows contour lines of the analytical tunneling probability \(\mathscr{T}_{\min}\) as a function of \(N\) and \(\gamma\) for a fixed runtime \(T=200\). Complementary numerical simulations (Fig.~\ref{fig:combined}b) illustrate the time evolution of the instantaneous ground-state fidelity \(F(s)= \bra{\psi_{g}(s)}\rho(s)\ket{\psi_{g}(s)}\) for three protocols. The Landau–Zener schedule exhibits a significant drop near the avoided crossing, indicating irreversible tunneling out of the ground state. The Roland-Cerf and dephasing optimal protocols also experience fidelity dips, but these are smaller and partially recovered by the end of the evolution, with the dephasing optimal schedule achieving the largest final fidelity. A detailed computational analysis of scaling follows in the next section. 

\subsubsection{Computational Scaling Analysis}

The adiabatic formulation of the Grover search fixes the scaling of the minimum spectral gap, \(g_{\min} = 1/\sqrt{N}\), but leaves open how the dephasing rate \(\gamma\) should be chosen to scale with the system size \(N\). To address this question, we introduce the dimensionless parameter \(\alpha := \gamma\sqrt{N}\) and analyze the asymptotic behavior of the minimal runtime \(T_{\min}\) from Eq.~\eqref{eq:Tmin} in different dephasing regimes. We note that a closely related scaling analysis was previously carried out by Avron and Fraas~\cite{avron}, who estimated the runtime by evaluating the characteristic timescale \(\tau\) in Eq.~\eqref{eq:optimal-speed}, using the estimate $\tau = \mathcal{O}\left( \frac{M(1/2)}{N} \right)$.

We note that for extremely low noise, $\gamma \ll \varepsilon = 1/T$, the system effectively recovers the unitary scenario; as noted by Avron and Fraas~\cite{avron}, this limit lies outside the dissipative adiabatic framework, which requires the adiabatic parameter $\varepsilon$ to be the smallest energy scale, in particular $\varepsilon \ll \gamma$. Physically, the total evolution time must be much longer than the dephasing time for the bath to produce the dephasing effect assumed in the derivation. Therefore, Eq.~\ref{eq:Tmin} is a nonzero-dephasing asymptotic formula and should not
be expected to converge to the Roland–Cerf runtime as \(\gamma\to0^+\). Indeed, its formal zero-dephasing limit gives $T_{\min} \to  0$, while $T_{\mathrm{RC}}$ remains finite; this signals the
breakdown of the dissipative expansion rather than a physical speedup.

\medskip
\textit{Weak dephasing} (\(\alpha \ll 1\)).
In the regime \(\varepsilon \ll \gamma \ll g_{\min}\), we use the above mentioned condition which directly leads to $T_{\min}\gg \frac{1}{\gamma} \gg O(\sqrt{N})$. This scaling confirms that weak
dephasing is computationally suboptimal but consistent with the Grover bound.

\begin{figure}[!t]
  \centering
  \includegraphics[width=1\columnwidth]{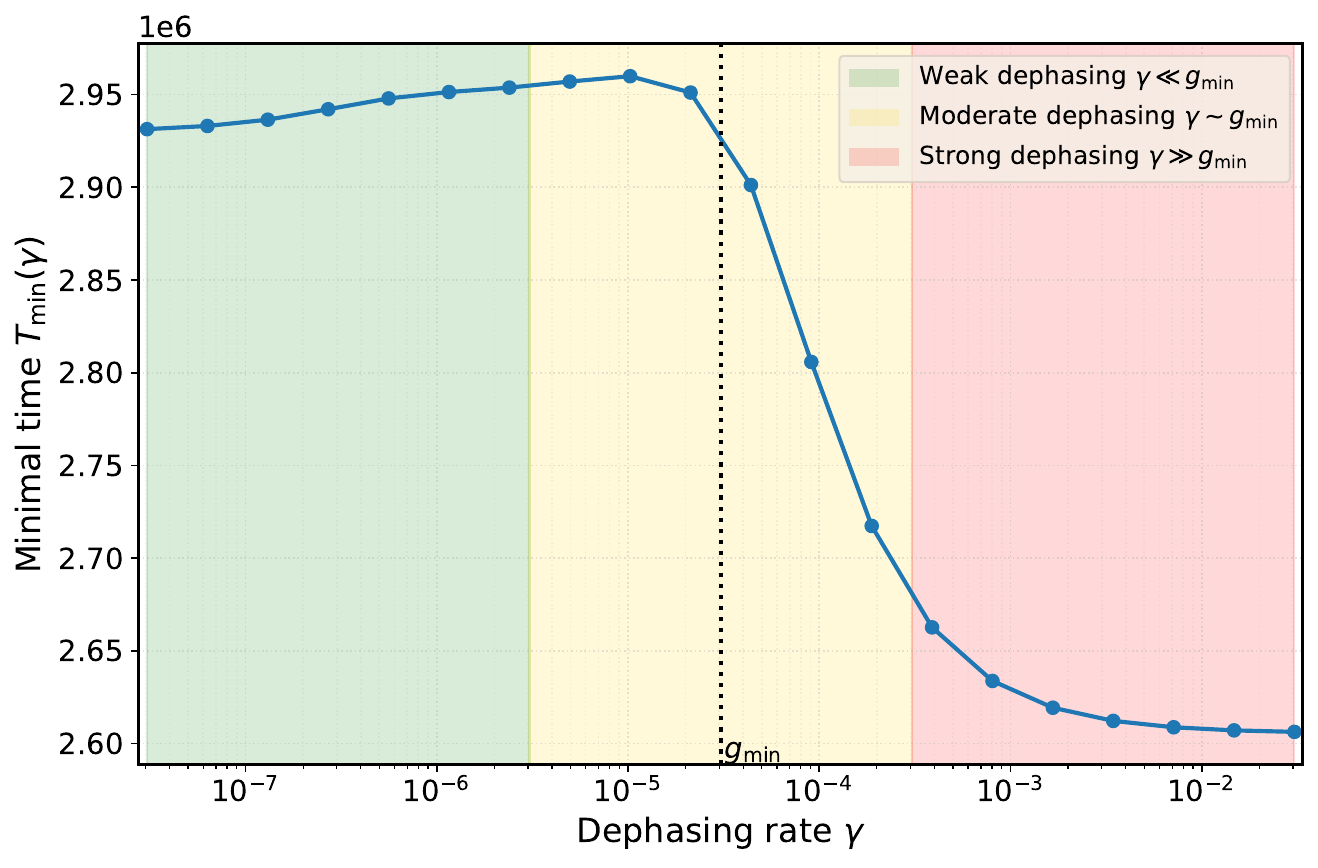}
  \caption{Minimal runtime $T_{\min}$ to reach $90\%$ fidelity for $n=30$ qubits as a function of $\gamma$. Shaded regions denote approximate weak (green), moderate (yellow), and strong (red) dephasing regimes.}
  \label{fig:Tmin_gamma}
\end{figure}

\medskip
\textit{Moderate dephasing} (\(\alpha = O(1)\)).
When the dephasing rate scales with the minimum gap, \(\gamma = \alpha/\sqrt{N}\) with \(\alpha = O(1)\), the bracketed factor in Eq.~\eqref{eq:Tmin} becomes the constant
\(c(\alpha) = \arctan(\alpha^{-1}) + \arctan(\alpha/2)\),
leading to
\begin{equation}
T_{\min} \simeq
\frac{2c(\alpha)^2}{\alpha\mathscr{T}}\sqrt{N}
= O(\sqrt{N}),
\label{eq:Tmin_matched}
\end{equation}
thus recovering the quadratic Grover speedup. As illustrated in Fig.~\ref{fig:Tmin_gamma}, this regime represents the maximum of the minimal runtime -- indicating that while the scaling with $N$ is optimal, the absolute prefactor is at its peak. The physical origin of this peak lies in the resonance between the environment and the system's characteristic energy splitting. Much like a laser tuned precisely to the transition frequency of an atom, dephasing in this regime is most effective at driving transitions out of the ground state. Because the noise is "tuned" to the minimum gap $g_{\min}$, it maximally induces leakage into the excited state, requiring the longest possible evolution time to satisfy the adiabatic condition and prevent population loss.

\medskip
\textit{Strong dephasing} (\(\alpha \gg 1\)).
For large dephasing, a leading-order expansion of Eq.~\eqref{eq:Tmin} shows that the sum of arctangent terms again tends to \(\pi/2\), yielding
\begin{equation}
T_{\min} \simeq
\frac{\pi^2}{2\gamma\mathscr{T}}
\!\left[1 + O(\alpha^{-1}) + O(N^{-1/2})\right]
= O(1/\gamma).
\label{eq:Tmin_strong}
\end{equation}
In this regime, the runtime becomes independent of $N$ and decreases inversely with $\gamma$, explaining the rapid decay on the right side of Fig.~\ref{fig:Tmin_gamma}. A caveat is that this asymptotic expression is still obtained from the dissipative adiabatic expansion, and must therefore be used only when the resulting time satisfies $\varepsilon = 1/T_{\mathrm{min}} \ll \gamma $. Remarkably, if the dephasing scales as $\gamma \sim N^{-a/2}$, the runtime follows $T_{\min} = O(N^{a/2})$, outperforming Grover’s $\sqrt{N}$ scaling whenever $a < 1$.

\begin{figure*}[!ht]
  \centering
  \includegraphics[width=\textwidth, height=0.35\textheight]{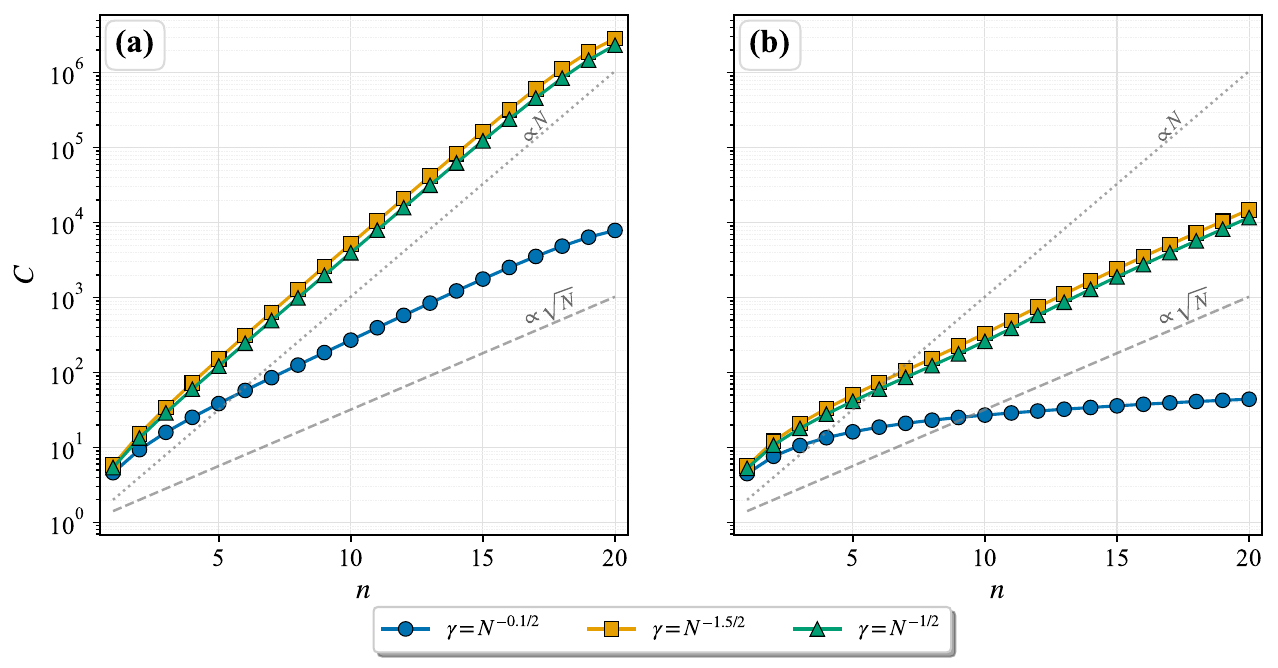}
  \caption{Scaling of the adiabatic quantity \(C\) from Eq.~\eqref{eq:adiabaticTheoremC} with qubit number \(n\) for dephasing \(\gamma\propto N^{-a/2}\) (blue \(a=0.1\), green \(a=1\), orange \(a=1.5\)). 
  \textbf{(a)} Linear schedule \(q(s)=s\): \(C\sim N\) (weak/matched) and \(C\sim\sqrt N\) (strong dephasing). 
    \textbf{(b)} Optimized schedule \(q_{\rm opt}(s)\): \(C\sim\sqrt N\) (weak/moderate) and \(C\sim\mathrm{const}\) (strong). 
    Grey lines show \(N\) and \(\sqrt N\).
    }
  \label{fig:adiabatic}
\end{figure*}

To confirm this behavior, we evaluate the adiabatic quantity $C$ (Eq.~\eqref{eq:adiabaticTheoremC}) entering the open-system adiabatic theorem by Venuti et al.\cite{Venuti2016Adiabaticity}, introduced previously. While this framework has mainly been studied in the context of general dissipative dynamics and steady-state preparation~\cite{zanardi2014coherent, venuti2017relaxation}, it has rarely been applied to concrete quantum algorithms. In particular, its implications for adiabatic quantum search in the presence of dephasing have not been explored. As shown in Fig.~\ref{fig:adiabatic}b, strong dephasing (\(\gamma \gg N^{-1/2}\)) drives \(C\)  to an \(N\)-independent plateau, so that \(T = O(1)\) suffices for any \(N\). This should be understood as a Zeno-type stabilization of the target eigenspace by a tailored dephasing mechanism, rather than as a conventional algorithmic speedup for an unstructured search. At first sight, this appears to contradict the optimality proof for Grover search time in the presence of a universal bath (see the Appendix of Roland and Cerf~\cite{rolandcerf}). The resolution lies in the physical limits on how large \(\gamma\) can be made for a \emph{universal} bath, as we now explain. 

\subsubsection{Maximal Dephasing Rate Under Time–Energy Constraints}

Pure dephasing in the energy eigenbasis -- where each off-diagonal element of the density matrix decays exponentially with rate \(\gamma\) -- can be viewed as a sequence of weak measurements of the projectors \(P_k = |E_k\rangle \langle E_k|\)~\cite{QuantumSearchMeasurement}. Using Kraus operators
\[
M_0 = \sqrt{1 - \gamma \Delta t} \, \mathds{1} , \quad M_k = \sqrt{\gamma \Delta t} \, P_k,
\]
one obtains 
\[
[\rho(t+\Delta t)]_{ij} = \begin{cases}
\rho_{ii}(t), & i = j, \\[6pt]
(1 - \gamma \Delta t) \, \rho_{ij}(t), & i \neq j,
\end{cases}
\]
which reproduces the exponential decay in the limit \(\gamma \Delta t \ll 1\). Thus \(\gamma\) is precisely the “measurement rate” of the system Hamiltonian \(H_S\). In the high‐dephasing regime \(\gamma \gg g_{\min}\) one would need
\(
  \Delta t \sim \frac{1}{\gamma} \;\ll\; \frac{1}{g_{\min}}
\)
to suppress coherences on the scale of the minimum gap \(g_{\min}\).  Yet the time–energy uncertainty principle forbids an \textit{unknown} energy splitting \(\Delta E\) from being resolved in any shorter time than
\(
  \Delta t \;\gtrsim\; \frac{\hbar}{\Delta E}.
\)
In a two‐level subspace the maximum possible spread is \(\Delta E_{\max} = g/2\) (see Remark \ref{remark:uncertainty}), so resolving the smallest gap demands
\(
  \Delta t \gtrsim \frac{\hbar}{g/2}
  \gtrsim \frac{\hbar}{g_{\min}},
\)
up to factors of order unity. 

A universal bath is defined by the interaction Hamiltonian
\[ H_{\mathrm{int}} = O_S \otimes B_{\mathrm{bath}},\]
where $O_S$ is an operator acting on the system Hilbert space and $B_{\mathrm{bath}}$ is an operator acting on the bath degrees of freedom. $O_S$ is chosen independently of the system Hamiltonian $H_S$, so the bath has no prior information about the system energy eigenbasis. The resulting noise therefore generally mixes different energy eigenspaces.

By contrast, a system‐specific bath couples as
\[H_{\mathrm{int}} \propto H_S \otimes B_{\mathrm{bath}},
  \qquad [H_{\mathrm{int}},H_S] = 0,\]
Because $H_{\mathrm{int}}$ commutes with $H_S$, the interaction is diagonal in the energy eigenbasis and does not induce transitions between energy levels. It therefore generates pure dephasing, which is operationally equivalent to continuous measurements of the energy projectors $P_k$. Since strengthening this coupling only increases the dephasing without changing the measurement basis, the corresponding rate $\gamma$ can be taken arbitrarily large, formally approaching the limit $\gamma \to \infty$ of infinitely frequent measurements. This leads to the \textit{quantum Zeno effect}, where sufficiently frequent measurements suppress coherent transitions and effectively freeze the state in its instantaneous eigenstate, without violating uncertainty bounds because the measurement basis is already known. Hence, no universal reservoir can induce dephasing faster than \(\hbar / g_{\min}\) for an unknown Hamiltonian. It can only approach this bound by effectively encoding system-specific information.

\medskip

To quantify the maximal admissible rate consistent with the Mandelstam–Tamm bound~\cite{MandelstamTamm1945}, which lower-bounds the time required for a state with a given energy uncertainty to evolve to an orthogonal state, we impose for each measurement interval:
\begin{equation}
\Delta t \ge \frac{\pi \hbar}{2\Delta E},
\label{eq:MT_bound_short}
\end{equation}
and require that the discrete measurement model reproduce the exponential decay with tolerance \(\delta \ll 1\):
\(
\big| e^{-\gamma \Delta t} - (1 - \gamma \Delta t) \big| \le \delta.
\)
Setting \(x = \gamma \Delta t\), the equality condition \(e^{-x} = 1 - x + \delta\) yields
\[
x = 1 + \delta + W_0(-e^{-(1+\delta)}),
\]
where \(W_0\) is the principal branch of the Lambert function. Expanding for small \(\delta\),
\(x \simeq \sqrt{2\delta} + \tfrac{2}{3}\delta\). Note that $x \le 1$ is required for the measurement channel to remain CPTP. Substituting the minimal interval \(\Delta t_{\min} = \pi\hbar/g\) gives the maximal dephasing rate
\begin{equation}
\gamma_{\max} \simeq \frac{g}{\pi \hbar}\!\left(\sqrt{2\delta}+\tfrac{2}{3}\delta\right), \qquad 0<\delta<1.
\label{eq:gamma_max_paper}
\end{equation}

\noindent
Any rate \(\gamma>\gamma_{\max}\) would require measurement intervals shorter than the uncertainty limit, effectively inducing rapid projective resets -- a genuine quantum‐Zeno freeze with runtime \(\mathcal{O}(1)\).

\begin{figure}[!t]
  \centering
  \includegraphics[width=1\columnwidth]{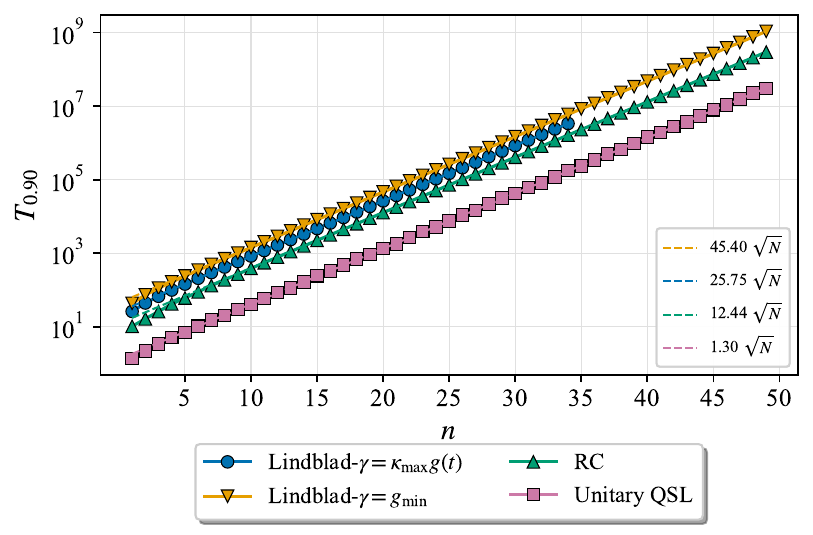}
  \caption{Minimum simulation time to reach 90\% fidelity under dephasing Lindblad evolution (Eq.~\eqref{eq:dephasing}) as a function of qubit number \(n\) for $\gamma = g_{\min}$ and $\gamma = \kappa_{\max} g(t)$ (Eq.~\eqref{eq:gamma_max_paper}, $\delta=e^{-1}$, other constants set to one), compared with unitary Roland–Cerf dynamics. The unitary QSL from Eq.~\eqref{eq:DL_QSL} is shown for reference. Dashed lines are color-matched $\sqrt{N}$ fits.}
\label{fig:last_fig}
\end{figure}
%width=0.85\textwidth, height=0.35\textheight

For the adiabatic Grover problem, where the avoided-crossing gap scales as \(g_{\min} \propto N^{-1/2}\), this analysis imposes a global upper bound \(\gamma_{\max}^{(\rm global)} = \mathcal O(N^{-1/2})\). Choosing \(\gamma = \mathcal O(g_{\min})\) keeps the system adiabatic, preserving the quadratic scaling \(T = \Theta(\sqrt{N})\): dephasing is strong enough to suppress diabatic transitions yet weak enough to avoid Zeno freezing. In practice, the best trade-off is achieved by letting the dissipator track the instantaneous gap, \(\gamma(s) = \kappa\, g(s)\), with \(\kappa \lesssim \kappa_{\max}\). This “gap-tracking” strategy saturates the local upper bound while ensuring global consistency with the time–energy constraint, maintaining Grover-like scaling across all system sizes.

Numerical results in Fig.~\ref{fig:last_fig} confirm that gap-tracking dephasing reduces the simulation time compared to constant dephasing, although the unitary protocol with the RC schedule remains faster. Here we chose $\delta = e^{-1}$ in Eq.~\eqref{eq:gamma_max_paper}, which corresponds to $x = 1$, the maximal value allowed for the measurement channel to remain CPTP. In the same figure, we also confirm numerically that the RC schedule does not saturate the unitary quantum speed limit, as discussed previously. Thus, while dephasing can aid the dynamics, it only surpasses unitary-optimized approaches within the Zeno regime, which is relevant primarily when the solution is already known and the focus is on state freezing rather than search speedup.

So far, we have derived a maximal dephasing bound consistent with time–energy constraints for a universal bath. It is natural to compare this with fundamental limits on evolution imposed by QSLs, which bound the minimal time required to evolve between states for a \emph{given} dynamical generator. A representative example is the Deffner–Lutz bound~\cite{Deffner2013},
\begin{equation}
\label{eq:DL_QSL} 
    \tau_{\rm QSL} \;\ge\; \frac{\sin^2\!\big(\mathcal L(\rho_0,\rho_\tau)\big)}{\displaystyle \frac{1}{\tau}\int_0^\tau dt\,\|\mathcal L_t(\rho_t)\|_1},
\end{equation}
where $\mathcal L_t$ is the Lindblad generator and $\mathcal L(\rho_0,\rho_\tau)$ the Bures angle between initial and final states.  
For pure dephasing, QSLs are tight and scale inversely with the dephasing rate, $\tau_{\rm QSL} = \mathcal O(1/\gamma)$~\cite{Mai2023}, meaning stronger dephasing increases the instantaneous speed allowed by the bound without constraining the Lindblad rates themselves. By contrast, Eq.~\eqref{eq:gamma_max_paper} constrains the \emph{admissible generator} for a universal bath lacking prior spectral knowledge. In this sense, our bound is a consistency condition on Lindbladians in adiabatic and algorithmic settings, whereas open-system QSLs apply once the generator is fixed. QSLs remain valid for system-specific or engineered reservoirs, while Eq.~\eqref{eq:gamma_max_paper} identifies when such generators cease to be universal.
% -----------------------------------------
% VI. CONCLUSION
% -----------------------------------------
\section{Conclusion}
\label{sec:conclusion}
In this paper, we revisited adiabatic quantum search in the presence of dephasing within the Lindbladian framework introduced by Avron et al. \cite{avron}, and applied it to the Grover-type Hamiltonian. Using their optimal-schedule construction, we derived explicit expressions for the time schedule, minimum runtime, and minimum achievable fidelity. We supported these results with numerical simulations. Our analysis shows that when dephasing acts along the instantaneous energy eigenbasis, populations remain confined to the adiabatic ground-state manifold while coherences between energy levels decay. The dynamics are determined primarily by the interpolation schedule and the dephasing rate~$\gamma$.

Three distinct dynamical regimes arise. When $\gamma \ll g_{\min}$, the bath’s effect is negligible and the system exhibits closed-system-like linear-time dynamics without speedup. When $\gamma \sim g_{\min}$, the system retains Grover’s quadratic speedup $T=\mathcal{O}(\sqrt{N})$ despite the presence of decoherence. In the opposite limit, $\gamma \gg g_{\min}$, continuous environmental monitoring induces a quantum‐Zeno effect that can freeze the dynamics, effectively reducing the runtime to constant order.

The apparent conflict between Zeno-enhanced acceleration and Grover’s optimal $\mathcal{O}(\sqrt{N})$ bound is resolved by recognizing that an arbitrarily fast Markovian bath cannot be universal: the Lindblad operators must implicitly encode information about the system Hamiltonian. This insight aligns with the time–energy uncertainty principle, which restricts the monitoring rate of an unknown Hamiltonian to the spectral gap, while prior knowledge lifts this constraint, enabling Zeno freezing.

Overall, we find that dephasing only outperforms unitary-optimized approaches in the Zeno regime, which is primarily relevant when the solution is already known and the goal is to freeze the state rather than perform the search. Our analysis here focuses on the theoretical scaling of the simulation time, assuming idealized conditions. In practical implementations on quantum hardware, additional overheads--such as control errors, finite temperature effects, and non-Markovian noise--must be taken into account, which can alter the effective runtime and achievable fidelity.

\section{Acknowledgments}
This work is based on a master’s thesis carried out at and financially supported by Siemens AG in collaboration with the Technical University of Munich (TUM). The authors thank Dr. Sarah Braun for insightful discussions.
% -----------------------------------------
% Appendices
% -----------------------------------------
\appendix
\appendix
\section{Derivation of the Dephasing Lindbladian}
\label{app:gksl}

We outline here the derivation of the general dephasing Lindbladian starting from the GKSL form \eqref{eq:Lindblad}.  
Let \(\{P_a\}_{a=1}^d\) denote the spectral projectors of \(H\), satisfying \(P_a P_b = \delta_{ab} P_a\) and \(\sum_a P_a = \mathds{1}\).  
To model an energy-preserving bath, we choose Lindblad operators that are diagonal in the energy basis:
\begin{equation}
L_j = \sum_a \sqrt{\tilde\gamma_{a j}}\, P_a,
\qquad
L_j^{\dagger} = \sum_b \sqrt{\tilde\gamma_{j b}}\, P_b,
\end{equation}
where the non-negative coefficients \(\tilde\gamma_{a j}\) specify the coupling strength of each level to bath channel \(j\).
Substituting this ansatz into the dissipative part of Eq.~\eqref{eq:Lindblad}:
\[
\mathcal{D}(\rho)
= 2\sum_j L_j\rho L_j^{\dagger}
- \sum_j \{L_j^{\dagger}L_j,\rho\},
\]
we compute each term separately. For the first contribution,
\begin{align}
2\sum_j L_j\rho L_j^{\dagger}
&= 2 \sum_{j,a,b} \sqrt{\tilde\gamma_{a j}\tilde\gamma_{j b}}\, P_a\rho P_b \notag\\
&= 2\sum_{a,b}\gamma_{ab} P_a\rho P_b,
\end{align}
where the rate matrix
\(
\gamma_{ab} := \sum_j \sqrt{\tilde\gamma_{a j}\tilde\gamma_{j b}}
\)
is Hermitian and positive semidefinite, as it is a Gram matrix of the square-root coefficients.
For the second and third contributions, note that for each \(j\),
\[
L_j^{\dagger}L_j
= \sum_{b,a}\sqrt{\tilde\gamma_{j b}\tilde\gamma_{a j}} P_bP_a
= \sum_a \sqrt{\tilde\gamma_{j a}\tilde\gamma_{a j}}\, P_a,
\]
so that
\begin{equation}
-\sum_j\{L_j^{\dagger}L_j,\rho\}
= -\sum_a \gamma_{aa}\{P_a,\rho\}.
\end{equation}
Combining the results above yields the general dephasing Lindbladian:
\begin{equation}
\label{eq:app_general_depha}
\mathcal{L}(\rho)
= -i[H, \rho]
+ 2\sum_{a,b}\gamma_{ab} P_a\rho P_b
- \sum_a \gamma_{aa}\{P_a,\rho\}.
\end{equation}
This generator is CPTP because \(\gamma_{ab}\ge0\).  
Moreover, since all \(L_j\) commute with \(H\), the average energy is conserved, \(\operatorname{tr}[H\,\mathcal{L}(\rho)] = 0\).
When all energy levels dephase uniformly, \(\gamma_{ab} = \gamma\,\delta_{ab}\), Eq.~\eqref{eq:app_general_depha} simplifies to
\begin{equation}
    \mathcal{L}(\rho)= -i[H,\rho] + 2\gamma \sum_a P_a\rho P_a
- \gamma\sum_a\{P_a,\rho\}
\end{equation}
Using the completeness relation $\sum_a P_a = \mathds{1}
$, it follows that $\sum_a P_a \rho = \rho$ and $\sum_a \rho P_a = \rho$. The anticommutator thus contributes $2\gamma \rho$ and cancels an equal term already present in the sum over $P_a \rho P_a$, since $\rho = \sum_{a,b} P_a \rho P_b = \sum_{a \neq b} P_a \rho P_b + \sum_a P_a \rho P_a$. Defining $\rho_{\text{off}} := \sum_{a \neq b} P_a \rho P_b$ for the off-diagonal (coherence) part and replacing $2\gamma \rightarrow \gamma$, we obtain
\[\mathcal{L}(\rho) = -i[H, \rho] - \gamma \sum_{a \neq b} P_a \rho P_b.\]

\section{Proof of Theorem~\ref{thm:AF}}
The original statement of this result was presented in Ref.~\cite{avron}, where only a sketch of the proof was provided. For completeness and to ensure full mathematical transparency, we give here a detailed and self-contained derivation following the same line of reasoning. We assume throughout that \(q(s)\) is a \(C^2\) monotonic schedule and that the spectrum of the instantaneous Hamiltonian \(H_q = \sum_j E_j(q) P_j(q)\) remains nondegenerate.

\noindent
Recall that for any linear map (superoperator) \(\mathcal{L}: B(H) \to B(H)\), its adjoint \(\mathcal{L}^*\) is defined with respect to the Hilbert–Schmidt inner product \(\langle A, B \rangle = \operatorname{tr}(A^\dagger B)\), i.e.
\begin{equation}
\label{eq:adjoint_trace2}
\operatorname{tr}(A^\dagger \mathcal{L}(B)) = \operatorname{tr}((\mathcal{L}^*(A))^\dagger B)
\quad \text{for all } A,B.
\end{equation}

For the purely dephasing Lindbladian \(\mathcal{L}_q\) associated with \(H_q\), each spectral projector \(P_j(q)\) satisfies
\[
\mathcal{L}_q(P_j(q)) = 0, \qquad
\mathcal{L}_q^*(P_j(q)) = 0, \quad \forall j.
\]

Let \(q = q(s)\) be a \(C^1\) schedule and let \(\rho(s) = \rho_{q,\varepsilon}(s)\) solve the master equation \eqref{eq:Lindbladian}. Define the ground-state population:
\[
p(s) = \operatorname{tr}[P_0(q(s))\,\rho(s)].
\]
Applying the chain rule gives:
\begin{equation}
\label{eq:pop_deriv}
\dot{p}(s) = \operatorname{tr}[P_0'(q)\,\rho(s)]\,\dot{q}(s)
+ \operatorname{tr}[P_0(q)\,\dot{\rho}(s)].
\end{equation}

Using \(\mathcal{L}_q^*(P_0(q)) = 0\), the Lindblad equation and \eqref{eq:adjoint_trace2} imply:
\[
\operatorname{tr}[P_0(q)\,\dot{\rho}(s)]
= \frac{1}{\varepsilon}\operatorname{tr}[P_0(q)\,\mathcal{L}_q(\rho(s))]
= \frac{1}{\varepsilon}\operatorname{tr}[\mathcal{L}_q^*(P_0(q))\,\rho(s)]
= 0.
\]
Thus, Eq.~\eqref{eq:pop_deriv} simplifies to:
\begin{equation}
\frac{d}{ds} \operatorname{tr}[P_0(q)\,\rho(s)]
= \operatorname{tr}[P_0'(q)\,\rho(s)]\,\dot{q}(s).
\end{equation}

We now switch to the Heisenberg picture, where the generator reads:
\[
\mathcal{L}_q^*(A) = i[H_q, A] - \gamma(q) \sum_{j \neq k} P_j A P_k.
\]
When acting on off-diagonal blocks \(P_a A P_b\) with \(a \neq b\), the spectral decomposition implies:
\begin{align*}
\mathcal{L}_q^*(P_a A P_b)
&= -i(E_a - E_b) P_a A P_b - \gamma(q) P_a A P_b \\
&= [i(E_a - E_b) - \gamma] P_a A P_b.
\end{align*}

Differentiating \(P_0(q)^2 = P_0(q)\) shows \(P_a P_0'(q) P_a = 0\) for all \(a\). Therefore, define:
\[
X(q) = \sum_{a \neq b} \frac{P_a\,P_0'(q)\,P_b}{i(E_a - E_b) - \gamma}.
\]
Then:
\begin{align}
\label{eq:adointP}
\mathcal{L}_q^*(X(q))
&= \sum_{a \neq b} \frac{\mathcal{L}_q^*(P_a P_0'(q) P_b)}{i(E_a - E_b) - \gamma}
= \sum_{a \neq b} P_a P_0'(q) P_b
= P_0'(q).
\end{align}

Substitute this into the population derivative:
\begin{align}
\label{eq:deriv_tr}
\frac{d}{ds} \operatorname{tr}[P_0(q)\,\rho(s)]
&= \operatorname{tr}[\mathcal{L}_q^*(X)\,\rho(s)]\,\dot{q}(s) \nonumber \\
&= \operatorname{tr}[X\,\mathcal{L}_q(\rho(s))]\,\dot{q}(s)
= \varepsilon\, \operatorname{tr}[X\,\dot{\rho}(s)]\,\dot{q}(s),
\end{align}
where we again used \eqref{eq:adjoint_trace2} and the master equation.

Now evaluate the tunneling rate at the final time:
\[
\mathscr{T}_{q, \varepsilon}(1) = 1 - \operatorname{tr}\bigl[P_0 \cdot \rho_{q, \varepsilon}(1)\bigr]
= - \int_{0}^{1} \frac{d}{ds} \operatorname{tr}[P_0(q)\,\rho(s)]\,ds.
\]
Using Eq.~\eqref{eq:deriv_tr}, we obtain:
\begin{equation}
\label{eq:tunneling}
\mathscr{T}_{q, \varepsilon}(1)
= -\varepsilon \int_{0}^{1} \operatorname{tr}[X\,\dot{\rho}(s)]\,\dot{q}(s)\,ds.
\end{equation}

Define \(F(s) = \operatorname{tr}[X(q(s))\,\rho(s)]\). Differentiating and integrating by parts gives:
\begin{align}
\label{eq:tunneling_long}
\mathscr{T}_{q,\varepsilon}(1)
&= -\varepsilon \int_{0}^{1} \dot{q} \,\dot{F}\, ds
+ \varepsilon \int_{0}^{1} \dot{q}^2\, \operatorname{tr}[X'(q)\rho]\, ds \nonumber \\
&= -\varepsilon \bigl[\dot{q} F\bigr]_{0}^{1}
+ \varepsilon \int_{0}^{1} \ddot{q} \, F \, ds
+ \varepsilon \int_{0}^{1} \dot{q}^2\, \operatorname{tr}[X'(q)\rho]\, ds.
\end{align}

The adiabatic theorem for dephasing Lindbladians implies \(\rho(s) = P_0(q(s)) + O(\varepsilon)\), so:
\[
F(s) = \operatorname{tr}[X P_0] + O(\varepsilon).
\]
Differentiate \(P_0^2 = P_0\) to obtain \(P_0 P_0' + P_0' P_0 = P_0'\). Sandwiching with \(P_a(\cdot)P_b\) and using orthogonality \(P_a P_b = \delta_{ab}P_a\) shows:
\[
P_a P_0' P_b = 0 \quad \text{unless exactly one of } a,b \text{ equals } 0.
\]
Thus, the non-vanishing blocks of \(P_0'\) connect the ground space to the excited subspace, allowing us to write:
\begin{align*}
    X &= \sum_{a \neq 0} \left(
\alpha_a\,P_a P_0' P_0 + \alpha_a^*\,P_0 P_0' P_a
\right),\\
\alpha_a &= \frac{i(E_a - E_0) + \gamma}{(E_a - E_0)^2 + \gamma^2}.
\end{align*}
Using \(P_a P_0 = 0\) for \(a \neq 0\), we get:
\[
X P_0 = \sum_{a \neq 0} \alpha_a P_a P_0' P_0
\quad \Rightarrow \;
\operatorname{tr}[X P_0] = 0 \; \Rightarrow \;
F(s) = O(\varepsilon).
\]

Insert \(\rho = P_0 + O(\varepsilon)\) and \(F = O(\varepsilon)\) into \eqref{eq:tunneling_long}. The boundary term and the \(\ddot{q}\) integral both become \(O(\varepsilon^2)\), so we obtain:
\begin{equation}
\label{eq:tunneling_tr}
\mathscr{T}_{q,\varepsilon}(1) =
\varepsilon \int_0^1 \dot{q}^2(s)\,
\operatorname{tr}[X'(q(s))\,P_0(q(s))]\, ds
+ O(\varepsilon^2).
\end{equation}

Finally, evaluate the trace:
\[
-\operatorname{tr}[X' P_0]
= 2\gamma(q) \sum_{a \neq 0}
\frac{\operatorname{tr}[P_a\, (P_0')^2]}
     {(E_a - E_0)^2 + \gamma^2(q)}
= 2M(q),
\]
so substituting into Eq.~\eqref{eq:tunneling_tr} yields the tunneling formula announced in Theorem~\ref{thm:AF}.

\section{Comparison of the Roland--Cerf time and the Mandelstam--Tamm QSL}
\label{app:QSLproof}

In this appendix, we show that the Roland--Cerf adiabatic schedule~\cite{rolandcerf} does not saturate the geometric QSL along the Grover adiabatic path.  
We do this by comparing the minimal evolution time along the path, given by the Mandelstam--Tamm bound, with the actual runtime of the Roland--Cerf schedule.

For the nondegenerate instantaneous ground state $|g(s)\rangle$ of the adiabatic Grover Hamiltonian $H(s)$, the Fubini--Study line element along the path satisfies
\begin{equation}
d\ell = \frac{ds}{g(s)},
\end{equation}
where $g(s)=E_1(s)-E_0(s)$ is the instantaneous spectral gap. This result follows from the standard eigenstate derivative identity~\cite{sakurai,rigolin2008}, together with the fact that for the adiabatic Grover Hamiltonian \(\partial_s H = H_p - H_0\) is constant, and \(|\langle e|\partial_s H|g\rangle|=1\) in the large-\(N\) limit.

The instantaneous velocity of state evolution in projective Hilbert space is given by the Anandan–Aharonov kinematic expression~\cite{AnandanAharonov1990} \(v_{\mathrm{AA}} = d\ell/dt = \Delta E(t)\), where the instantaneous energy uncertainty is
\[
\Delta E(t) = \sqrt{\langle \psi(t)|H^2|\psi(t)\rangle - \langle \psi(t)|H|\psi(t)\rangle^2}.
\]
For a two-level system, \(\Delta E_{\max} = g(s)/2\), achieved when ground and excited populations are equal (cf. Remark \ref{remark:uncertainty}). The Mandelstam–Tamm bound~\cite{MandelstamTamm1945,DeffnerCampbell2017} thus gives the minimal evolution time along a fixed path:
\begin{equation}
T_{\mathrm{QSL}} = \int_0^1 \frac{d\ell}{v_{\max}(s)} = \int_0^1 \frac{d\ell}{\Delta E_{\max}(s)} = 2 \int_0^1 \frac{ds}{g^2(s).}
\label{eq:TQSL}
\end{equation}
By contrast, the Roland–Cerf local adiabatic rule \(\dot{s} = \varepsilon g^2(s)\) (see \cite{rolandcerf}, equation 17) implies
\begin{equation}
T_{\mathrm{RC}} = \frac{1}{\varepsilon}\int_0^1 \frac{ds}{g^2(s)}.
\end{equation}

Since $T_{\mathrm{RC}} = \frac{1}{2\varepsilon}\,T_{\mathrm{QSL}}$, equality would formally require $\varepsilon = 1/2$.  
However, adiabaticity requires $0<\varepsilon\ll 1$ \cite{rolandcerf}, and moreover the QSL in Eq.~\eqref{eq:TQSL} is overly optimistic: it assumes the state saturates the maximal energy uncertainty at all times, while the actual adiabatic state remains almost entirely in the ground state manifold for most of the evolution except near the minimum gap. A more realistic QSL along the adiabatic path would therefore be \(T_{\mathrm{QSL}}^{\rm actual} = \int_0^1 \frac{d\ell}{\Delta E(s)}\,,\) which is larger than the bound in Eq.~\eqref{eq:TQSL}. Consequently, although the Roland–Cerf schedule achieves the optimal asymptotic scaling, its actual runtime remains slower than the true QSL for all \(N\).

\medskip
\begin{remark}
\label{remark:uncertainty}
In a two-level system with instantaneous Hamiltonian
\(H'=-\frac{g}{2}\,\sigma_z,\) any pure state \(|\psi\rangle\) with ground-state population
\(p = |\langle 0|\psi\rangle|^2\) has energy variance
\[
(\Delta E)^2
= \bigl\langle H'^2\bigr\rangle - \langle H'\rangle^2
= g^2\,p(1-p)\;\le\;\frac{g^2}{4},
\]
so the maximal energy spread \(\Delta E_{\max} = g/2\) is attained at \(p=\tfrac12\).
\end{remark}

\section{Derivation of the adiabatic schedule}
\label{app:grover-schedule}
\noindent
To derive the explicit form of the optimal schedule, we symmetrize the interpolation parameter around the minimum gap by writing
\[q(s) = \tfrac{1}{2} + u(s), \qquad -\tfrac{1}{2} \leq u \leq \tfrac{1}{2}, \quad 0 \leq s \leq 1,\]
so that the avoided crossing is centered at $u = 0$. For notational convenience, we set $\ell := N-1$ and $A := N\gamma^2$. For the Grover two-level Hamiltonian, the gap and the square root of the mass term in terms of $u$ are
\begin{equation*}
\label{eq:g2_M}
g^2(u) = \frac{1 + 4\ell u^2}{N}, \qquad
\sqrt{M(u)} = \frac{\sqrt{\gamma\,\ell N}}{(1+\ell u^2)\;\sqrt{1+\ell u^2 + A}}.
\end{equation*}
The optimal schedule is determined by integrating the ODE from Eq.~\eqref{eq:optimal-speed}, which leads to
\[
I(u) = \int_{-1/2}^{u(s)} \sqrt{M(u)}\, du = \tau\, s,
\]
\textbf{Step 1: Evaluation of $I(u)$.} Pulling out the constant
\(C:=\sqrt{\gamma\ell N}\), setting \(x=2\sqrt{\ell}\,u\) and defining $A :=N\gamma^2$ yields
\begin{align*}
I(u) &= C \int_{-\gamma/2}^{u(s)}
   \frac{du}{(1 + 4\ell u^2)\sqrt{1 + 4\ell u^2 + N\gamma^2}} \\
     &= \frac{\sqrt{\gamma N}}{2} 
   \int_{-\sqrt{\ell}}^{x(s)}
   \frac{dx}{(1 + x^2)\sqrt{1 + x^2 + A}}.
\end{align*}
we then use the substitution $x = \tan\phi$, so $dx = \sec^2\phi\, d\phi$ and $1 + x^2 = \sec^2\phi$. This gives
\(\sqrt{1 + x^2 + A} = \sec\phi\, \sqrt{1 + A\cos^2\phi},\) and the integral becomes
\[I = \int \frac{\cos\phi\, d\phi}{\sqrt{1 + A\cos^2\phi}}.\]
Next, set $y = \sin\phi$, so $dy = \cos\phi\, d\phi$ and $\cos^2\phi = 1 - y^2$, yielding
\begin{align*}
I &= \int \frac{dy}{\sqrt{1 + A(1 - y^{2})}}
     = \int \frac{dy}{\sqrt{a - b y^{2}}} \notag \\
  &= \frac{1}{\sqrt{a}}\, \arcsin\!\left( \sqrt{\frac{b}{a}}\, y \right) + C,
\end{align*}
with a := 1 + A,\; b := A. Alternatively, in terms of $x$, we have
\[
J(x) = \frac{1}{\sqrt{1 + A}} \,
\arcsin\left(
  \frac{x \sqrt{A}}{\sqrt{(1 + A)(1 + x^2)}}
\right) + C.
\]
The definite integral is therefore
\begin{equation}
\label{eq:definite}
    I(u)=J\bigl(x(s)\bigr)-J(-\sqrt{\ell}),
\end{equation}
and the \textbf{constant} \textbf{$\tau$} reads  
\begin{equation}
\label{eq:constant}
\begin{split}
\tau &= J(\sqrt{\ell}) - J(-\sqrt{\ell}) \\
     &= \frac{2}{\sqrt{1 + A}} \arcsin\left( \frac{\gamma \sqrt{\ell}}{\sqrt{1 + A}} \right) \\
     &= 2\Theta_0, \quad \Theta_0 := \sin^{-1}\left( \frac{\gamma \sqrt{N - 1}}{\sqrt{1 + N\gamma^2}} \right).
\end{split}
\end{equation}
\textbf{Step 2: Relating $s$ and $u$.}
Dividing Eq.~\eqref{eq:definite} by the constant in Eq.~\eqref{eq:constant} defines a strictly monotonic map \( u \mapsto s \), given by:
\begin{equation}
\label{eq:su}
s(u) = \frac{1}{2} + \frac{1}{2\Theta_0} 
\arcsin\left( \frac{x \sqrt{A}}{\sqrt{(1 + A)(1 + x^2)}} \right),
\end{equation}
where \( x = 2\sqrt{\ell} u \).

\vspace{0.3em}
\noindent
\textbf{Step 3: Inverting $s(u)$ to obtain $u(s)$.}
Equation~\eqref{eq:su} can be rearranged as
\begin{equation*}
    \arcsin\!\left(
  \frac{2u\,\sqrt{N(N-1)\,\gamma^2}}
       {\sqrt{(1+N\gamma^2)\,[1+4(N-1)u^2]}}
\right)
= 2\Theta_0\!\left(s-\tfrac12\right).
\end{equation*}

\noindent
\begin{samepage}
Letting $\theta(s):=(2s-1)\Theta_0$ and taking the sine of both sides, we
ask \texttt{SymPy}'s \verb|solve| routine to isolate $u$, obtaining

\begin{equation}
\label{eq:u_s}
u(s)=\frac12\,
\frac{\sin \theta(s)\,\sqrt{1+N\gamma^2}}
{\sqrt{(N-1)\bigl[N\gamma^2-
(1+N\gamma^2)\sin^2\theta(s)\bigr]}}.
\end{equation}
\end{samepage}

\noindent
Finally, reverting to $q=\tfrac12+u$ yields the explicit optimal schedule
quoted in the main text, Eq.~\eqref{eq:optimal_schedule}.

% -----------------------------------------
% Bibliography
% -----------------------------------------
\bibliographystyle{apsrev4-2}  
\bibliography{literature} 

@article{avron,
  author    = {J. E. Avron and M. Fraas and G. M. Graf},
  title     = {Optimal time schedule for adiabatic evolution},
  journal   = {Phys. Rev. A},
  volume    = {82},
  number    = {4},
  pages     = {040304(R)},
  year      = {2010},
  doi       = {10.1103/PhysRevA.82.040304}}

@book{QMAhard,
  title = {Classical and Quantum Computation},
  author = {Kitaev, Alexei Yu. and Shen, Alexander and Vyalyi, Mikhail N.},
  series = {Graduate Studies in Mathematics},
  volume = {47},
  publisher = {AMS},
  year = {2002},
  isbn = {9780821832295},
  url = {https://bookstore.ams.org/gsm-47}
}

@article{farhi,
  author    = {Edward Farhi and Jeffrey Goldstone and Sam Gutmann and Michael Sipser},
  title     = {Quantum Computation by Adiabatic Evolution},
  journal   = {Science},
  volume    = {292},
  number    = {5516},
  pages     = {472--476},
  year      = {2001},
  note      = {arXiv:quant-ph/0001106},
  url       = {https://arxiv.org/abs/quant-ph/0001106}
}

@inproceedings{equivalent,
  title = {Adiabatic Quantum Computation Is Equivalent to Standard Quantum Computation},
  author = {Aharonov, Dorit and van Dam, Wim and Kempe, Julia and Landau, Zeph and Lloyd, Seth and Regev, Oded},
  booktitle = {Proceedings of the 45th Annual IEEE Symposium on Foundations of Computer Science (FOCS)},
  pages = {42--51},
  year = {2004},
  publisher = {IEEE},
  doi = {10.1109/FOCS.2004.8},
  url = {https://arxiv.org/abs/quant-ph/0405098}
}

@article{shortcuts,
  author    = {D. Guéry-Odelin and A. Ruschhaupt and A. Amunduzain and J. G. Muga},
  title     = {Shortcuts to Adiabaticity: Concepts, Methods, and Applications},
  journal   = {Reviews of Modern Physics},
  volume    = {91},
  number    = {4},
  pages     = {045001},
  year      = {2019},
  doi       = {10.1103/RevModPhys.91.045001},
  url       = {https://link.aps.org/doi/10.1103/RevModPhys.91.045001}
}

@article{rolandcerf,
  author    = {Jérémie Roland and Nicolas J. Cerf},
  title     = {Quantum Search by Local Adiabatic Evolution},
  journal   = {Physical Review A},
  volume    = {65},
  number    = {4},
  pages     = {042308},
  year      = {2002},
  doi       = {10.1103/PhysRevA.65.042308},
  eprint    = {quant-ph/0107015},
  archivePrefix = {arXiv},
  primaryClass = {quant-ph},
  url       = {https://arxiv.org/abs/quant-ph/0107015}
}

@article{grover,
  title = {Quantum Mechanics Helps in Searching for a Needle in a Haystack},
  author = {Grover, Lov K.},
  journal = {Phys. Rev. Lett.},
  volume = {79},
  issue = {2},
  pages = {325--328},
  numpages = {0},
  year = {1997},
  month = {Jul},
  publisher = {American Physical Society},
  doi = {10.1103/PhysRevLett.79.325},
  url = {https://link.aps.org/doi/10.1103/PhysRevLett.79.325}
}

@article{12,
  title = {Robustness of adiabatic quantum computation},
  author = {Childs, Andrew M. and Farhi, Edward and Preskill, John},
  journal = {Phys. Rev. A},
  volume = {65},
  issue = {1},
  pages = {012322},
  numpages = {10},
  year = {2001},
  month = {Dec},
  publisher = {American Physical Society},
  doi = {10.1103/PhysRevA.65.012322},
  url = {https://link.aps.org/doi/10.1103/PhysRevA.65.012322}
}

@article{13,
author    = {M. Steffen and W. van Dam and T. Hogg and G. Breyta and I. Chuang},
title     = {Experimental Implementation of an Adiabatic Quantum Optimization Algorithm},
journal   = {Physical Review Letters},
volume    = {90},
number    = {6},
pages     = {067903},
year      = {2003},
doi       = {10.1103/PhysRevLett.90.067903}
}

@article{albash2015decoherence,
  title = {Decoherence in adiabatic quantum computation},
  author = {Albash, Tameem and Lidar, Daniel A.},
  journal = {Phys. Rev. A},
  volume = {91},
  issue = {6},
  pages = {062320},
  numpages = {19},
  year = {2015},
  month = {Jun},
  publisher = {American Physical Society},
  doi = {10.1103/PhysRevA.91.062320},
  url = {https://link.aps.org/doi/10.1103/PhysRevA.91.062320}
}

@article{14,
  title = {Thermally Assisted Adiabatic Quantum Computation},
  author = {Amin, M. H. S. and Love, Peter J. and Truncik, C. J. S.},
  journal = {Phys. Rev. Lett.},
  volume = {100},
  issue = {6},
  pages = {060503},
  numpages = {4},
  year = {2008},
  month = {Feb},
  publisher = {American Physical Society},
  doi = {10.1103/PhysRevLett.100.060503},
  url = {https://link.aps.org/doi/10.1103/PhysRevLett.100.060503}
}

@article{weakcoupling,
  author    = {Tameem Albash and Sergio Boixo and Daniel A. Lidar and Paolo Zanardi},
  title     = {Quantum adiabatic Markovian master equations},
  journal   = {New Journal of Physics},
  volume    = {14},
  pages     = {123016},
  year      = {2012},
  doi       = {10.1088/1367-2630/14/12/123016}
}

@article{Farhi1998,
  title = {Analog analogue of a digital quantum computation},
  author = {Farhi, Edward and Gutmann, Sam},
  journal = {Phys. Rev. A},
  volume = {57},
  issue = {4},
  pages = {2403--2406},
  numpages = {0},
  year = {1998},
  month = {Apr},
  publisher = {American Physical Society},
  doi = {10.1103/PhysRevA.57.2403},
  url = {https://link.aps.org/doi/10.1103/PhysRevA.57.2403}
}

@article{Demirplak2003,
  author    = {Mustafa Demirplak and Stuart A. Rice},
  title     = {Adiabatic population transfer with control fields},
  journal   = {The Journal of Physical Chemistry A},
  volume    = {107},
  number    = {46},
  pages     = {9937--9945},
  year      = {2003},
  doi       = {10.1021/jp030708a}
}

@article{Berry2009,
  author    = {Michael V. Berry},
  title     = {Transitionless quantum driving},
  journal   = {Journal of Physics A: Mathematical and Theoretical},
  volume    = {42},
  number    = {36},
  pages     = {365303},
  year      = {2009},
  doi       = {10.1088/1751-8113/42/36/365303}
}

@article{Berry84,
  author = {Berry, M. V.},
  title = {Quantal phase factors accompanying adiabatic changes},
  journal = {Proc. R. Soc. A},
  volume = {392},
  number = {1802},
  pages = {45--57},
  year = {1984},
  doi = {10.1098/rspa.1984.0023}
}

@article{Bason2012,
  author    = {Mark G. Bason and Michael Viteau and Nicola Malossi and Paul Huillery and Edvard Arimondo and Donatella Ciampini and Rosario Fazio and Vittorio Giovannetti and Rosario Mannella and Oliver Morsch},
  title     = {High-fidelity quantum driving},
  journal   = {Nature Physics},
  volume    = {8},
  number    = {2},
  pages     = {147--152},
  year      = {2012},
  doi       = {10.1038/nphys2170}
}

@article{QuantumSearchMeasurement,
  title = {Quantum search by measurement},
  author = {Childs, Andrew M. and Deotto, Enrico and Farhi, Edward and Goldstone, Jeffrey and Gutmann, Sam and Landahl, Andrew J.},
  journal = {Physical Review A},
  volume = {66},
  pages = {032314},
  year = {2002},
  doi = {10.1103/PhysRevA.66.032314},
  eprint = {quant-ph/0204013}
}

@article{AvronFraas,
  author    = {J. E. Avron and M. Fraas and G. M. Graf and P. Grech},
  title     = {Adiabatic Theorems for Generators of Contracting Evolutions},
  journal   = {Communications in Mathematical Physics},
  volume    = {314},
  pages     = {163--191},
  year      = {2012},
  doi       = {10.1007/s00220-012-1504-1}
}

@article{Venuti2016Adiabaticity,
  author    = {Lorenzo Campos Venuti and Tameem Albash and Daniel A. Lidar and Paolo Zanardi},
  title     = {Adiabaticity in open quantum systems},
  journal   = {Physical Review A},
  volume    = {93},
  number    = {3},
  pages     = {032118},
  year      = {2016},
  publisher = {American Physical Society},
  doi       = {10.1103/PhysRevA.93.032118}
}

@book{NielsenChuang2000,
  title     = {Quantum Computation and Quantum Information},
  author    = {Nielsen, Michael A. and Chuang, Isaac L.},
  year      = {2000},
  edition   = {1st},
  publisher = {Cambridge University Press},
  address   = {Cambridge, UK},
  isbn      = {978-0-521-63503-5},
  url       = {https://books.google.com/books/about/Quantum_Computation_and_Quantum_Informat.html?id=65FqEKQOfP8C},
  note      = {10th Anniversary Edition: 2010}
}

@article{landauzener,
  title     = {Non-adiabatic crossing of energy levels},
  author    = {Clarence Zener},
  journal   = {Proceedings of the Royal Society of London. Series A, Containing Papers of a Mathematical and Physical Character},
  volume    = {137},
  number    = {833},
  pages     = {696--702},
  year      = {1932},
  doi       = {10.1098/rspa.1932.0165},
  url       = {https://royalsocietypublishing.org/doi/10.1098/rspa.1932.0165}
}

@article{trajectories,
  title   = {Adiabatic quantum trajectories in engineered reservoirs},
  author  = {King, Emma C. and Giannelli, Luigi and Menu, Rapha{\"e}l and Kriel, Johannes N. and Morigi, Giovanna},
  journal = {Quantum},
  volume  = {8},
  pages   = {1428},
  year    = {2024},
  month   = jul,
  doi     = {10.22331/q-2024-07-30-1428},
  url     = {https://doi.org/10.22331/q-2024-07-30-1428},
  issn    = {2521-327X}
}

@article{Schuetzhold2006,
  title = {Adiabatic quantum algorithms as quantum phase transitions: First versus second order},
  author = {Sch\"utzhold, Ralf and Schaller, Gernot},
  journal = {Phys. Rev. A},
  volume = {74},
  issue = {6},
  pages = {060304},
  numpages = {4},
  year = {2006},
  month = {Dec},
  publisher = {American Physical Society},
  doi = {10.1103/PhysRevA.74.060304},
  url = {https://link.aps.org/doi/10.1103/PhysRevA.74.060304}
}

@article{AdiabaticResponse,
  author    = {J. E. Avron and M. Fraas and G. M. Graf},
  title     = {Adiabatic Response for Lindblad Dynamics},
  journal   = {Journal of Statistical Physics},
  volume    = {148},
  pages     = {800--823},
  year      = {2012},
  doi       = {10.1007/s10955-012-0550-6}
}

@article{zeno,
  author    = {Jesse Berwald and Nick Chancellor and Raouf Dridi},
  title     = {Grover Speedup from Many Forms of the Zeno Effect},
  journal   = {Quantum},
  volume    = {8},
  pages     = {1532},
  year      = {2024},
  doi       = {10.22331/q-2024-11-20-1532},
  eprint    = {2305.11146},
  archivePrefix = {arXiv},
  primaryClass  = {quant-ph},
  url       = {https://arxiv.org/abs/2305.11146}
}

@article{Peter,
  title   = {Quantum Dissipative Search via Lindbladians},
  author  = {Eder, Peter J. and Fin{\v{z}}gar, Jernej Rudi and Braun, Sarah and Mendl, Christian B.},
  journal = {Physical Review A},
  volume  = {111},
  number  = {4},
  pages   = {042430},
  year    = {2025},
  doi     = {10.1103/PhysRevA.111.042430},
  url     = {https://doi.org/10.1103/PhysRevA.111.042430},
  publisher = {American Physical Society}
}

@article{brachistochrone,
  author    = {A. T. Rezakhani and D. A. Lidar and P. Zanardi},
  title     = {Quantum Adiabatic Brachistochrone},
  journal   = {Physical Review Letters},
  volume    = {103},
  number    = {8},
  pages     = {080502},
  year      = {2009},
  doi       = {10.1103/PhysRevLett.103.080502}}

@article{SarandyLidar2005,
  title = {Adiabatic approximation in open quantum systems},
  author = {Sarandy, M. S. and Lidar, D. A.},
  journal = {Physical Review A},
  volume = {71},
  number = {1},
  pages = {012331},
  year = {2005},
  doi = {10.1103/PhysRevA.71.012331}
}

@article{Facchi2002,
  title={Quantum Zeno Subspaces},
  author={Facchi, Paolo and Pascazio, Saverio},
  journal={Phys. Rev. Lett.},
  volume={89},
  number={8},
  pages={080401},
  year={2002},
  doi={10.1103/PhysRevLett.89.080401},
  eprint={quant-ph/0201115},
  archivePrefix={arXiv}
}

@article{Verstraete2009,
  title={Quantum computation and quantum-state engineering driven by dissipation},
  author={Verstraete, Frank and Wolf, Michael M. and Cirac, J. Ignacio},
  journal={Nature Physics},
  volume={5},
  number={9},
  pages={633--636},
  year={2009},
  doi={10.1038/nphys1342},
  url={https://doi.org/10.1038/nphys1342},
  publisher={Nature Publishing Group}
}

@article{AltshulerKroviRoland2010,
  title = {Anderson localization makes adiabatic quantum optimization fail},
  author = {Altshuler, Boris and Krovi, Hari and Roland, Jérémie},
  journal = {Proceedings of the National Academy of Sciences},
  volume = {107},
  number = {28},
  pages = {12446--12450},
  year = {2010},
  doi = {10.1073/pnas.1002116107}
}

@article{stateart,
  title = {Adiabatic quantum computation},
  author = {Albash, Tameem and Lidar, Daniel A.},
  journal = {Rev. Mod. Phys.},
  volume = {90},
  issue = {1},
  pages = {015002},
  numpages = {64},
  year = {2018},
  month = {Jan},
  publisher = {American Physical Society},
  doi = {10.1103/RevModPhys.90.015002},
  url = {https://link.aps.org/doi/10.1103/RevModPhys.90.015002}
}

@inproceedings{vanDamMoscaVazirani2001,
  author    = {Wim van Dam and Michele Mosca and Umesh V. Vazirani},
  title     = {How Powerful Is Adiabatic Quantum Computation?},
  booktitle = {Proceedings of the 42nd Annual IEEE Symposium on Foundations of Computer Science (FOCS)},
  pages     = {279--287},
  year      = {2001},
  doi       = {10.1109/SFCS.2001.959902},
  url       = {https://arxiv.org/abs/quant-ph/0206003}
}

@article{rigolin2008,
  author = {Rigolin, G. and Ortiz, G. and Ponce, V. H.},
  title = {Beyond the quantum adiabatic approximation: Adiabatic perturbation theory},
  journal = {Phys. Rev. A},
  volume = {78},
  pages = {052508},
  year = {2008},
  doi = {10.1103/PhysRevA.78.052508},
  url = {https://doi.org/10.1103/PhysRevA.78.052508}
}

@book{sakurai,
  author = {Sakurai, J. J. and Napolitano, J.},
  title = {Modern Quantum Mechanics},
  publisher = {Cambridge University Press},
  year = {2017},
  isbn = {9781108473224},
  url = {https://www.cambridge.org/core/books/modern-quantum-mechanics/8D1B15C5EFC2E6FBCF1C57F65F3B4572}
}

@article{MandelstamTamm1945,
  author  = {Mandelstam, L. and Tamm, I.},
  title   = {The uncertainty relation between energy and time in non-relativistic quantum mechanics},
  journal = {Journal of Physics (USSR)},
  volume  = {9},
  pages   = {249--254},
  year    = {1945}
}

@article{AnandanAharonov1990,
  author  = {Anandan, J. and Aharonov, Y.},
  title   = {Geometry of quantum evolution},
  journal = {Physical Review Letters},
  volume  = {65},
  number  = {13},
  pages   = {1697--1700},
  year    = {1990},
  doi     = {10.1103/PhysRevLett.65.1697}
}

@article{DeffnerCampbell2017,
  author  = {Deffner, S. and Campbell, S.},
  title   = {Quantum speed limits: from Heisenberg’s uncertainty principle to optimal quantum control},
  journal = {Journal of Physics A: Mathematical and Theoretical},
  volume  = {50},
  number  = {45},
  pages   = {453001},
  year    = {2017},
  doi     = {10.1088/1751-8121/aa86c6}
}

@article{Mai2023,
  title={Tight and attainable quantum speed limit for open systems},
  author={Mai, Zi-yi and Yu, Chang-shui},
    journal={Physical Review A},
    volume={108},
    number={5},
    pages={052207},
    year={2023},
    publisher={American Physical Society},
    doi={10.1103/PhysRevA.108.052207},
    eprint={2309.10308},
    archivePrefix={arXiv},
    primaryClass={quant-ph}
    }

@article{Deffner2013,
    title={Quantum speed limit for non-Markovian dynamics},
    author={Deffner, Sebastian and Lutz, Eric},
    journal={Physical Review Letters},
    volume={111},
    number={1},
    pages={010402},
    year={2013},
    publisher={American Physical Society},
    doi={10.1103/PhysRevLett.111.010402},
    eprint={1302.5069},
    archivePrefix={arXiv},
    primaryClass={quant-ph}
    }

@article{venuti2017relaxation,
  author       = {Venuti, Lorenzo Campos and Albash, Tameem and Marvian, Milad and Lidar, Daniel A. and Zanardi, Paolo},
  title        = {Relaxation versus adiabatic quantum steady-state preparation},
  journal      = {Physical Review A},
  volume       = {95},
  number       = {4},
  pages        = {042302},
  year         = {2017},
  publisher    = {American Physical Society},
  doi          = {10.1103/PhysRevA.95.042302}
}

@article{zanardi2014coherent,
  author       = {Zanardi, Paolo and Venuti, Lorenzo Campos},
  title        = {Coherent quantum dynamics in steady-state manifolds of strongly dissipative systems},
  journal      = {Physical Review Letters},
  volume       = {113},
  number       = {24},
  pages        = {240406},
  year         = {2014},
  publisher    = {American Physical Society},
  doi          = {10.1103/PhysRevLett.113.240406}
}
\end{document}